\documentclass[aps,pra,twocolumn]{revtex4-1}

\usepackage{amsmath}
\usepackage{amssymb}
\usepackage{epsfig}
\usepackage{color}
\usepackage{caption}
\usepackage{subcaption}
\usepackage{bbm}	
\usepackage{adjustbox}
\usepackage{float}
\usepackage{verbatim}
\usepackage[bookmarks]{hyperref}  

\newcommand* {\bra}[1]{\ensuremath{\langle {#1} |}}
\newcommand* {\ket}[1]{\ensuremath{| {#1} \rangle}}

\begin{document}
	
	\title{Range of applicability of the Hu-Paz-Zhang master equation}

	\author{G. Homa}
	\email{ggg.maxwell1@gmail.com}
	\affiliation{Department of Physics of Complex Systems, E\"{o}tv\"{o}s Lor\'and University, ELTE, P\'azm\'any P\'eter s\'et\'any 1/A, H-1117 Budapest, Hungary}

	\author{A. Csord\'as}
	\email{csordas@tristan.elte.com}
	\affiliation{Department of Physics of Complex Systems, E\"{o}tv\"{o}s Lor\'and University, ELTE, P\'azm\'any P\'eter s\'et\'any 1/A, H-1117 Budapest, Hungary}
	
	\author{M. A. Csirik}
	\email{csirik.mihaly@wigner.mta.hu}
	\affiliation{Hylleraas Centre for Quantum Molecular Sciences, Department of Chemistry,
		University of Oslo, P.O. Box 1033 Blindern, N-0315 Oslo, Norway﻿}
	\affiliation{Institute for Solid State Physics and Optics, Wigner Research Centre, Hungarian Academy of Sciences, 
		P.O. Box 49, H-1525 Budapest, Hungary}
	
	\author{J. Z. Bern\'ad}
	\email{zsolt.bernad@um.edu.mt}
	\affiliation{Department of Physics, University of Malta, Msida MSD 2080, Malta}
	\affiliation{Institut f\"{u}r Angewandte Physik, Technische Universit\"{a}t Darmstadt, D-64289 Darmstadt, Germany}
	
	\date{\today}
	
	\begin{abstract}
		We investigate a case of the Hu-Paz-Zhang master equation of the Caldeira-Leggett model without Lindblad
		form obtained in the weak-coupling limit up to the second-order perturbation. In our study, we use Gaussian
		initial states to be able to employ a sufficient and necessary condition, which can expose positivity violations
		of the density operator during the time evolution. We demonstrate that the evolution of the non-Markovian
		master equation has problems when the stationary solution is not a positive operator, i.e., does not have physical
		interpretation. We also show that solutions always remain physical for small-times of evolution. Moreover, we
		identify a strong anomalous behavior, when the trace of the solution is diverging. We also provide results for the
		corresponding Markovian master equation and show that positivity violations occur for various types of initial
		conditions even when the stationary solution is a positive operator. Based on our numerical results, we conclude
		that this non-Markovian master equation is superior to the corresponding Markovian one.
		
	\end{abstract}

	\maketitle

	\section{Introduction}
	\label{I}
	
	A density operator completely describes the state of a quantum mechanical system and it is defined as a positive trace class operator of trace one \cite{Neumann}. A quantum system in study 
	can be subject to interactions with its environment, which is colloquially referred to as
	an open quantum system. It is expected that the whole system evolves unitarily and, by tracing out the environment's 
	degrees of freedom, one obtains a positive trace preserving map acting on the states of the open system \cite{Davies}. If one further assumes an
	initially uncorrelated joint state, then a stronger kind of positivity, called complete positivity, is obtained \cite{Kraus}. Some particular aspects of this assumption have been discussed in Refs. \cite{Pechukas,Hakim,Romero,Buzek,Salgado}. 
	In physical applications, these maps are subject to further approximations, which either leads to Markovian or non-Markovian master equations \cite{book1}. However, the positivity of the approximation-free map 
	may be violated by various approximations implying that complete positivity fails as well. 
	
	A known case is the Caldeira-Leggett model \cite{CL} of the quantum Brownian motion \cite{Grabert,Weiss}, 
	where different approaches may result in a master equation, which  may not preserve the positivity of the density operator for short times \cite{Ambegaokar,Hu_Paz,Diosi,Gnutzmann}. In the model of 
	Unruh and Zurek \cite{Unruh} (where the environment is modeled differently from the Caldeira-Leggett model) issues have also been found with respect to rapid decoherence for short time evolutions.
	The well-known master equation of Caldeira and Leggett has been extended by Hu, Paz, and Zhang (HPZ), who obtained an exact non-Markovian master equation \cite{Hu_Paz},
	\begin{eqnarray}
	&&i \hbar \frac{\partial \hat{\rho}}{ \partial t}=\left [\hat{H}_0,\hat{\rho}\right]-i D_{pp}(t)[\hat{x},[\hat{x},\hat{\rho}]]
	\nonumber \\
	&& \qquad + \lambda(t) [\hat{x},\{\hat{p},\hat{\rho}\}]+2 i D_{px}(t)[\hat{x},[\hat{p},\hat{\rho}]], \nonumber
	\end{eqnarray}
	where $\hat{H}_0$ is the Hamiltonian of the open quantum system. $D_{pp}(t)$, $\lambda(t)$ and $D_{px}(t)$ are time-dependent coefficients for which one has explicit expressions 
	(see \cite{Hu_Paz} or \cite{yuhal}). A particular case of this master equation, when the interaction between the system and environment is weak, is given by Eq. \eqref{diffcoeff} for the explicit expressions 
	of the time-dependent coefficients. This case covers both the Caldeira-Leggett master equation \cite{CL} of high temperatures and an extension for lower temperatures 
	\cite{CLLT}. Despite the weak-coupling approximation the master equation has still found applications even decades later in several areas of quantum mechanics, such as quantum optomechanics \cite{Eisert} or quantum 
	estimation theory \cite{Paris}. These works consider the perturbative approach in the weak-coupling up to the second-order, which is also the first non-vanishing term in the perturbation series \cite{book1}. 
	In fact, this version of the HPZ master equation has drawn much attention in the last decade and therefore it is worth while to investigate, in detail, the circumstances under which the time evolution is able to preserve 
	the positivity of the density operator.
	
	The main parameters of the Caldeira-Leggett model are the temperature of the thermal bath and the spectral density of the environment. In the phenomenological 
	modeling, one expects that the spectral density goes to zero for very high frequencies. A special case is when the spectral density is proportional to the frequency for small values of frequency, i.e., the ohmic 
	spectral density, which gives rise to a frequency-independent damping rate. Other spectral densities have also been subject to investigations; see, e.g., \cite{Hu_Paz,Fleming,Garg}. In this paper, we choose the ohmic 
	spectral density with a Lorentz-Drude cutoff function. Furthermore, we consider the open quantum system to be a quantum harmonic oscillator.
	
	Recently, questions related to the positivity preservation of several Markovian master equations were investigated with the help of purities of density operators \cite{HBL}. 
	The authors exploited the fact that the purity indicates positivity violation when it takes values bigger than one. They have been able to identify cases where positivity violations occur.
	Unfortunately, the purity is a necessary but not sufficient condition to determine the positivity of a self-adjoint operator with trace one. In this paper, we consider a non-Markovian master equation 
	with its Markovian counterpart, which is obtained from the non-Markovian one by taking the limits in the coefficients  $t\rightarrow \infty$. Both the Markovian and the non-Markovian master equation 
	can be formally solved  \cite{Ford,Fleming,HBL} for all possible initial conditions. However, the obtained solutions in the phase-space representation 
	cannot determine, in general the positivity of associated Weyl operators,- \cite{Kastler,Nicola}, because one has to verify either a non-countable or a countable set of inequalities. In the special case of 
	Gaussian density operators, all the eigenvalues can be analytically determined \cite{Joos,BCSH}, and furthermore their structure implies that the Gaussian solution is 
	positive if and only if the purity is between zero and one. In particular, results in \cite{Fleming} imply that these types of master equations preserve the Gaussian form of 
	any initially Gaussian state for all times. Therefore, in the case of a Gaussian ansatz, we are able to use a necessary and sufficient condition to monitor the positivity of the evolving density operator.
	Furthermore, both master equations can be transformed into a system of ordinary differential equations.
	
	The paper is organized as follows. In Sec.~\ref{II}.~we introduce the non-Markovian master equation and derive the system of linear differential equations for coefficients of the Gaussian ansatz.  
	In Sec. \ref{III}, we study the positivity of the stationary solution. In parameter space we identify regions where positivity violations can occur. 
	Concrete examples of these violations are given in Sec.~\ref{IV}.
	Here, we concentrate on the differences of the Markovian and non-Markovian time evolutions of initial Gaussian density operators.  
	Section ~\ref{V} summarizes our main results. Technical details are provided in the three Appendices.
	
	\section{Non-Markovian master equation with Gaussian initial conditions}
	\label{II}
	
	In this section we discuss basic features of the HPZ master equation \cite{HR,Hu_Paz,Ford} by focusing on terms up to the second-order expansion in the weak-coupling strength 
	\cite{Breuer-Kappler}. The non-Markovian master equation for a quantum harmonic oscillator with physically observable frequency $\omega_p$ and mass $m$ reads
	\begin{eqnarray}
	&&i \hbar \frac{\partial \hat{\rho}}{ \partial t}=\left [\frac{\hat{p}^2}{2m}+\frac{m \omega^2_p(t) \hat{x}^2}{2},\hat{\rho}\right]-i D_{pp}(t)[\hat{x},[\hat{x},\hat{\rho}]]
	\nonumber \\
	&& \qquad + \lambda(t) [\hat{x},\{\hat{p},\hat{\rho}\}]+2 i D_{px}(t)[\hat{x},[\hat{p},\hat{\rho}]], \label{NMEQ}
	\end{eqnarray}
	where $[,]$ stands for commutators while $\{,\}$ for anti-commutators. In the weak-coupling limit the coefficients in the second-order expansion entering the master equation read   
	\begin{eqnarray}\label{diffcoeff}
	\omega^2_p(t)&=&\omega^2_b-\frac{2}{m}\int_{0}^t{ds D(s)\cos(\omega_0 s)},\nonumber \\
	\lambda(t)&=&\frac{1}{m \omega_0}\int_{0}^t{ds D(s)\sin(\omega_0 s)},\nonumber \\
	D_{px}(t)&=&\frac{1}{2 m \omega_0}\int_{0}^t{ds D_1(s)\sin(\omega_0 s)},\nonumber \\
	D_{pp}(t)&=&\int_{0}^t{ds D_1(s)\cos(\omega_0 s)},
	\end{eqnarray}
	where $\omega_b$ contains the environment-induced frequency shift of the original oscillator frequency $\omega_0$. We have introduced the following correlation functions:	
	\begin{eqnarray}
	D(s)&=&\int_0^{\infty}{d \omega J(\omega) \sin(\omega s)}, \label{eq:D(s)_def} \\
	D_1(s)&=&\int_0^{\infty}{d \omega J(\omega) \coth \left(\frac{\hbar \omega}{2 k_B T}\right) \cos(\omega s)}, \label{eq:D1(s)_def}.
	\end{eqnarray}
	where $T$ is the temperature of the thermal bath. Making use of an Ohmic spectral density with a Lorentz-Drude type function and a high-frequency cutoff $\Omega$,
	\begin{equation}	
	J(\omega)=\frac{2 m \gamma }{\pi} \omega \frac{\Omega^2}{\Omega^2+\omega^2}, \nonumber
	\end{equation}
	where $\gamma$ is the frequency-independent damping constant, the bath correlation  $D(s)$ can be determined analytically as
	\begin{equation}
	D(s)=2 m \gamma \Omega^2 \exp (-\Omega  s ), \quad s \ge 0.
	\end{equation}
	For the other correlation function $D_1(s)$ see Eq.~(\ref{eq:d1sumshort}) in Appendix \ref{Appuseful}. Furthermore, for $t>0$,
	\begin{eqnarray}
	\omega^2_p(t)&=&\omega^2_0+2 \gamma \Omega-\frac{2}{m}\int_{0}^t{ds D(s)\cos(\omega_0 s)} =\omega^2_0+2 \gamma \Omega \nonumber \\
	&-&\frac{2 \gamma  \Omega^2 }{\Omega^2+\omega^2_0} e^{-\Omega t } \left[\Omega  e^{\Omega t}-\Omega \cos \left(\omega_0 t\right)+\omega_0 \sin \left(\omega _0 t\right)\right], \nonumber 
	\end{eqnarray}
	where for $t \gg 1$, $\omega_p(t)$ is approximately equal to $\omega_0$, and 
	\begin{eqnarray}
	&&\lambda(t)= \nonumber \\
	&&=\frac{\gamma}{\omega_0}\frac{\Omega^2 }{\Omega^2+\omega^2_0} e^{-\Omega t } \left[\omega_0  e^{\Omega t}-\omega_0 \cos \left(\omega_0 t\right)-\Omega \sin \left(\omega _0 t\right)\right]. \nonumber 
	\end{eqnarray}
	Closed formulas for $D_{px}(t)$ and $D_{pp}(t)$ are given in Appendix \ref{Appuseful}. It is important to note that in the high-temperature limit 
	$k_B T \gg \hbar \Omega \gg \hbar \omega_0$, we have $\omega_p(t \to \infty)= \omega_0$, $D_{px}(t \to \infty)=\gamma k_B T/(\hbar \Omega)$, $D_{pp}(t \to \infty)= 2 m \gamma k_B T$, and $\lambda(t \to \infty)=\gamma$, 
	which yields exactly the Caldeira-Leggett master equation, i.e., the  term $ \gamma k_B T/(\hbar \Omega) [\hat{x},[\hat{p},\hat{\rho}]]$ is very small compared to the other two terms. Furthermore, 
	these coefficients also cover an extended master equation of Caldeira {\it et al.} \cite{CLLT} for lower temperatures by taking only 
	$\Omega \gg \omega_0$ in \eqref{diffcoeff}, which results in their finding $D_{pp}(t \to \infty)=m \gamma \hbar \omega_0 \coth{\hbar \omega_0/(2 k_B T)}$. However, in this particular case of the 
	HPZ master equation the weak damping assumption $\omega_0 \gg \gamma$ is not required.
	
	Now, we rewrite Eq. \eqref{NMEQ} in the position representation
	\begin{eqnarray}
	&&i \hbar \frac{\partial}{\partial t} \rho(x,y,t)=\Big[ \frac{\hbar^2 }{2m}\left(\frac{\partial^2}{\partial y^2}-
	\frac{\partial^2}{\partial x^2} \right) +\frac{m \omega^2_p(t)}{2} \left(x^2-y^2\right) \nonumber \\
	&& \qquad  -i D_{pp}(t) (x-y)^2 
	-i \hbar\lambda(t) (x-y) \left(\frac{\partial}{\partial x}-
	\frac{\partial}{\partial y} \right) \nonumber \\
	&& \qquad +2\hbar D_{px}(t) (x-y) \left(\frac{\partial}{\partial x}+
	\frac{\partial}{\partial y} \right) \Big]  \rho(x,y,t). \label{CROME}
	\end{eqnarray}
	
	Naively, the non-Markovian master equation starts at $t=0$ as a von Neumann equation, because all the time-dependent coefficients in \eqref{diffcoeff} are zero for $t=0$, except for $\omega_p(t)$. 
	This would imply that positivity violations never occur around $t=0$.  We prove this fact rigorously 
	for an arbitrary Gaussian initial state in Appendices \ref{sec:dpp_app} and \ref{sec:initial_jolt}. 
	For longer times it is not guaranteed that positivity will not be violated. Another property of \eqref{CROME} is that the Gaussian initial state remains Gaussian during the whole evolution. In \cite{Fleming}, 
	the time evolution of a Wigner function [see Eq. (78) of their paper] starting from an arbitrary initial condition is given. If this initial Wigner function 
	is Gaussian, then this result shows that at an arbitrary time $t>0$, the solution is also a Gaussian with time-dependent coefficients in the exponent. The Wigner function and $\rho(x,y,t)$ are connected by
	Wigner-Weyl transformation, which maps a Gaussian function to Gaussian ones. Consequently, if we choose  $\rho(x,y,t=0)$ to be Gaussian it will be Gaussian at later times too, 
	but with time-dependent coefficients. More concretely, we consider the following Gaussian in the position representation:
	\begin{eqnarray}
	&&\rho(x,y,t)=\exp\{-A (t)  \left( x-y \right)^2-iB(t)  \left( x^2-y^2 \right)  \nonumber \\
	&& ~ -C(t) \left( x+y \right) ^{2}
	-iD(t) (x-y)-E(t)(x+y) -N(t)\}, \label{xyform} \nonumber \\
	\end{eqnarray}
	where the time-dependent parameters $A$, $B$, $C$, $D$, $E$, and $N$ are real because $\hat{\rho}$ is self-adjoint. 
	Assuming positive $A(t)$ and $C(t)$ the eigenvalue problem in the position representation for a fixed $t$,
	\begin{equation}
	\int_{-\infty}^\infty \rho(x,y) \phi_n(y)\,dy=\lambda_n \phi_n(x)
	\end{equation}
	has been considered in detail in Ref. \cite{BCSH}. The spectrum $\{\lambda_n\}_{n \in \mathbb{N}_0}$ of \eqref{xyform}
	depends only on $A$ and $C$ for all $t\geqslant 0$:
	\begin{eqnarray}
	\lambda_n&=&\lambda_0 \lambda^n, \nonumber \\
	\lambda_0&=&\frac{2\sqrt{C}}{\sqrt{A}+\sqrt{C}},\quad \lambda=\frac{\sqrt{A}-\sqrt{C}}{\sqrt{A}+\sqrt{C}}. \nonumber
	\end{eqnarray}
	If $0<A<C$, then the Gaussian self-adjoint operator fails to be positive. 
	Clearly, all  eigenvalues are in the interval $[0,1]$ iff
	\begin{equation}
	A \geq C \geq 0.
	\label{good_spectrum}
	\end{equation}
	If Eq. \eqref{good_spectrum} is not true at a given time $t$, then the Gaussian function $\rho(x,y,t)$ has no physical interpretation, and Eq. \eqref{good_spectrum} is a sufficient and necessary 
	condition to detect unphysical behavior during the time evolution. 
	We are going to test its validity by investigating $A/C$. Note that the purity is given by $\textrm{Tr}\,\hat{\rho}^2=\sqrt{C/A}$. 
	
	The time-dependent coefficients $A$, $B$, $C$, $D$, and $E$ obey a system of nonlinear nonautonomous differential equations.  However, using the transformation 
	\begin{equation}
	\rho(k,\Delta,t)=\int_{-\infty}^\infty dx \, e^{ikx} \rho\Big( x+\frac{\Delta}{2},x-\frac{\Delta}{2},t\Big), \label{transform}
	\end{equation}
	given in \cite{Unruh,BCSH}, we obtain
	the equation of motion for $\rho(k,\Delta,t)$,
	\begin{eqnarray}
	\frac{\partial}{\partial t} \rho(k,\Delta,t)&=& \biggl(\frac{\hbar k}{m} \frac{\partial}{\partial \Delta} 
	-\frac{m \omega_p^2(t)}{\hbar} \Delta \frac{\partial}{\partial k}-\frac{D_{pp}(t)}{\hbar} \Delta^2 \biggr. \nonumber\\
	&& \biggl.-2\lambda(t) \Delta \frac{\partial}{\partial \Delta}-2D_{px}(t)k \Delta \biggr) \rho(k,\Delta,t). \nonumber
	\end{eqnarray}
	Note that the above equation of motion contains only first-order derivatives and therefore it is easier to construct its solutions. In this representation, the Gaussian form of \eqref{xyform}
	is also preserved and reads
	\begin{eqnarray}
	\rho(k,\Delta,t)&=&\exp\bigl\{-c_1(t)k^2-c_2(t)k \Delta -c_3(t) \Delta^2 \bigr.\nonumber\\
	&& \bigl.\quad \quad-ic_4(t) k-ic_5(t) \Delta-c_6(t)\bigr\}, \label{Ga}
	\end{eqnarray}
	where the time-dependent coefficients $c_1,c_2,c_3,c_4,c_5,$ and $c_6$ are real and obey the following system of \emph{linear} differential equations:
	\begin{eqnarray}
	\dot{c}_1&=&\frac{\hbar c_2}{m}, \quad \dot{c}_2=2D_{px}(t)+\frac{2\hbar c_3}{m}-2 \frac{m \omega_p^2(t)}{\hbar} c_1-2\lambda(t) c_2, \nonumber\\
	\dot{c}_3&=&\frac{D_{pp}(t)}{\hbar}-\frac{m \omega_p^2(t)}{\hbar} c_2-4\lambda(t) c_3,
	\quad \dot{c}_4=\frac{\hbar c_5}{m},  \nonumber\\
	\dot{c}_5&=&-\frac{m \omega_p^2(t)}{\hbar} c_4-2\lambda(t) c_5, \quad \dot{c}_6=0. \label{difftosolve}
	\end{eqnarray} The first three and the last three equations decouple.
	The first three can be written compactly as follows: 
	\begin{equation}
	\dot{\mathbf{c}}(t)=\mathbf{M}(t)\mathbf{c}(t)+\mathbf{v}(t),
	\label{eq:main_diffeq}
	\end{equation}
	where  $\mathbf{c}^T(t)=(c_1,c_2,c_3)$ (the superscript $T$ denotes the transposition),
	\begin{eqnarray}
	\mathbf{M}(t)=\begin{pmatrix}
	0 & \frac{\hbar}{m} & 0 \\
	-2 \frac{m \omega_p^2(t)}{\hbar}  & -2\lambda(t) & \frac{2\hbar}{m} \\
	0 & - \frac{m \omega_p^2(t)}{\hbar} & -4\lambda(t)
	\end{pmatrix}, \label{eq:m(t)_matrix}
	\end{eqnarray}
	and
	\begin{equation}
	\mathbf{v}(t)=\begin{pmatrix}
	0 \\
	2 D_{px}(t) \\
	D_{pp}(t)/\hbar
	\end{pmatrix}.\nonumber 
	\end{equation}
	The coefficients $A$, $B$, and $C$ are related to $\mathbf{c}$ through the transformation \eqref{transform} as
	\begin{equation}
	A=  c_3-\frac{c^2_2}{4c_1},   \hspace{0.5em} B=-\frac{c_2}{4c_1},  \hspace{0.5em}
	C=\frac{1}{16 c_1}.
	\label{eq:A_and_C_for_cs}
	\end{equation}
	We can already see the advantage of the new phase-space representation $\rho(k, \Delta)$ because solving \eqref{eq:main_diffeq} is
	better suited for our subsequent investigation of the ratio $A/C$. However, the solution of \eqref{eq:main_diffeq} is still not simple because the matrices 
	$\mathbf{M}(t)$ and $\mathbf{M}(t')$ do not commute at different times $t \neq t'$ and the vector $\mathbf{v}(t)$ is also time-dependent. A formal solution with the help
	of a time-ordered exponential can be given, but does not seem to be helpful for us. Therefore, we are going to focus on the numerical solutions of \eqref{eq:main_diffeq} and to carry out 
	a brief analysis on the stationary state.
	
	\section{A brief analytical study of the stationary state}
	\label{III}
	
	In this section, we investigate the positivity of the stationary state. After a long time, a Markovian limit is obtained, which yields
	\begin{eqnarray}\label{diffcoeffM}
	\omega^2_p(t \to \infty)&=&\left(\omega^{(M)}_p\right)^2, \quad \lambda(t\to\infty)=\lambda^{(M)},\nonumber \\
	D_{px}(t\to \infty)&=&D^{(M)}_{px},\quad D_{pp}(t\to \infty)=D^{(M)}_{pp},
	\end{eqnarray}
	where the details about Markovian values (denoted with superscripts $M$) are given in Appendix \ref{Appuseful}. Thus, 
	$\mathbf{M}(t)$ and $\mathbf{v}(t)$ tend to constants $\mathbf{M}^{(M)}$ and $\mathbf{v}^{(M)}$.  The stationary solution of $\mathbf{c}(t)$ can be expressed as:
	\begin{equation}
	\mathbf{c}^{(M)}=-[\mathbf{M}^{(M)}]^{-1}\mathbf{v}^{(M)}. \nonumber
	\end{equation}
	Approaching the stationary state is governed by the three eigenvalues of $\mathbf{M}$, which are $(-2) \lambda(t)$ and 
	$ (-2)\left(\lambda(t)\pm\sqrt{\lambda^2(t)-\omega_p^2(t)}\right) $. For $t>0$ real parts of all three eigenvalues are negative, and
	thus $\mathbf{M}(t)$ is contractive, which ensures that starting from arbitrary initial conditions $\mathbf{c}(0)$, the trajectory $\mathbf{c}(t)$ tends to its Markovian limit. 
	In the asymptotic region, $\lambda(t)$ and $\omega_p(t)$ must be replaced by their respective Markovian values.
	
	%%%%%%%%%%%%%%%%%%%%%%%%%%%%%%%%%%%%%%%%%%%%%%%%%%%%%%%%%%%%%%%%%%%%%%%%%%%%%%%%
	\begin{figure}[ht!]
		\includegraphics[angle=270,width=10cm,trim=0cm 3cm 0cm 0cm]{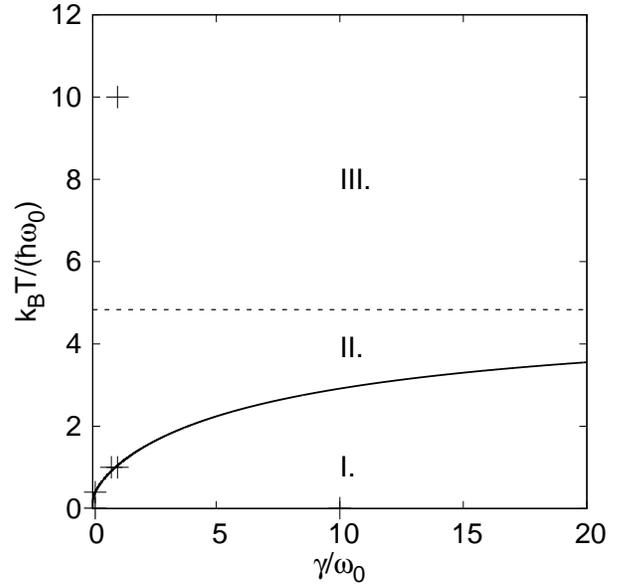}
		\caption{Parameter space plot of $k_B T/(\hbar\omega_0)$ vs $\gamma/\omega_0$ at fixed $\Omega/\omega_0=20$. 
			The solid thick line shows the critical line  $\gamma=\gamma_\mathrm{crit}(k_BT)$. The dashed line indicates the temperature $\tilde{T}$. 
			Regions I., II. and III. are discussed in the main text. Crosses indicate the parameters used in Figs.~\ref{case1}--\ref{case25}. \label{density}}. 
	\end{figure}
	%%%%%%%%%%%%%%%%%%%%%%%%%%%%%%%%%%%%%%%%%%%%%%%%%%%%%%%%%%%%%%%%%%%%%%%%%%%%%%%%
	
	In the asymptotic regime, where all the time-dependent coefficients of the equation of motion have already reached their stationary values, 
	the test $A/C \geq 1$ can be written as 
	\begin{equation}
	\frac{A^{(M)}}{C^{(M)}}=\frac{\left(D_{pp}^{(M)}\right)^2+4 m \lambda^{(M)}  D_{pp}^{(M)} D_{px}^{(M)}}{m^2\left(\lambda^{(M)}\right)^2  \left(\omega^{(M)}\right)^2}.
	\end{equation}
	On the critical line $A^{(M)}/C^{(M)}=1$ the damping factor $\gamma$ can be expressed as 
	\begin{equation}
	\gamma =\gamma_\mathrm{crit}(\Omega,k_B T,\omega_0)=\frac{\Omega^2+\omega_0^2}{\Omega}\cdot \frac{\coth^2 \left(\frac{\hbar\omega_0}{2k_BT}\right)-1}{Z(\Omega,k_BT,\omega_0)},
	\label{eq:gamma_crit_line}
	\end{equation}
	where
	\begin{eqnarray}
	&&Z(\Omega,k_BT,\omega_0)= 2-4\frac{k_BT}{\hbar\omega_0}\coth\frac{\hbar\omega_0}{2k_BT}\biggl[-1 
	+\frac{\hbar\Omega}{2\pi k_BT} \times \biggr.\nonumber \\
	&&\biggl. \times\biggl( \!\!  \Psi\left( \frac{i\hbar\omega_0}{2\pi k_B T} \right) 
	+\Psi\left( \frac{-i\hbar\omega_0}{2\pi k_B T} \right)-2\Psi\left( \frac{\hbar\Omega}{2\pi k_B T}\right) \!\!\biggr)\!\! \biggr] \label{eq:full_Z}
	\end{eqnarray}
	and $\Psi$ is the digamma function \cite{Stegun}. 
	The denominator $Z(\Omega,k_BT,\omega_0)$ has a zero if we vary $k_BT$, and  thus there exists a certain temperature $\tilde{T}$ at which the damping factor 
	$\gamma$ tends to infinity on the critical line; see Fig. \ref{density}. Clearly, above $\tilde{T}$, the stationary solution is a density operator for any damping factor 
	$\gamma$; see region III in Fig.~\ref{density}. The stationary solution is not a density operator in region I, i.e., $T<\tilde{T}$ and $\gamma> \gamma_\mathrm{crit}$. 
	In this parameter region we can choose any initial condition for which  the time evolution for $\hat{\rho}(t)$ eventually violates the positivity of the density operator. 
	Regions II and III of Fig.~\ref{density} guarantee that the asymptotic state is physically allowed,
	but this does not guarantee that the full time evolution is physical. We can also observe that very weak damping $\gamma \ll \omega_0$ allows us to chose the temperature $T$ 
	arbitrarily. This is in accordance with the result in Ref. \cite{CLLT}. However, a positive stationary solution still is not a guarantee for a meaningful time evolution, because issues
	might appear for several kind of
	initial conditions, especially if we choose the parameters of the master equations close to the critical line $\gamma_\mathrm{crit}$.
	
	Analytical approximations for the critical line can be made in two cases. If  $\Omega \gg \omega_0$ one can expect
	(see Fig.~\ref{density}) that $k_B\tilde{T}$ is on the $\hbar\Omega$ scale.   
	Let us  introduce the quantity $x= {k_B \tilde{T}}/({\hbar \Omega})$. If $\Omega \gg \omega_0$ looking for the zeros for $Z$ in Eq.\eqref{eq:full_Z} the leading terms  are 
	\begin{equation}
	0=\pi x + \gamma_\mathrm{EM} + \Psi\left( \frac{1}{2 \pi x} \right) ,
	\label{eq:eq_for_univ_number}
	\end{equation}
	where $\gamma_\mathrm{EM}$ is the Euler-Mascheroni constant, which is approximately $0.577$. 
	Solving \eqref{eq:eq_for_univ_number} for $x$, one gets  $k_B  \tilde{T}\approx 0.240395 \cdot \hbar \Omega$ for large $\Omega$. 
	It should be noted that this result has been previously found by Ref. \cite{Lampo2}, where the stationary state has been investigated from the point of view of the Heisenberg uncertainty principle. In the case
	of Gaussian density operators the Heisenberg uncertainty principle and our test condition $A/C\geqslant1$ are the same constraints on the parameter space of the master equation.
	
	\begin{figure*}[ht!]
		\begin{subfigure}{.48\textwidth}
			\includegraphics[angle=270,width=9.0cm]{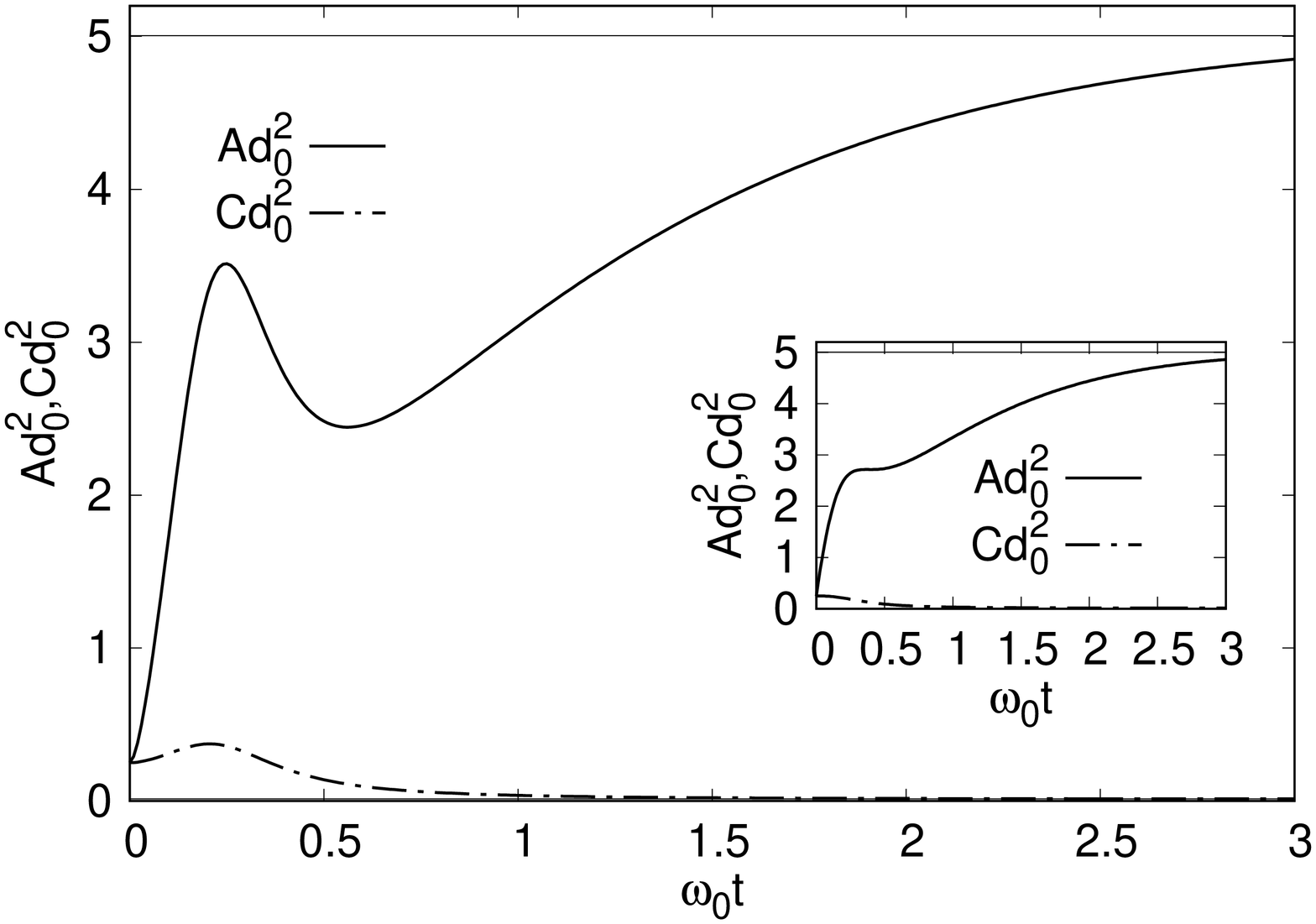}
			\caption{}
		\end{subfigure}\hspace{1em}
		\begin{subfigure}{.48\textwidth}
			\includegraphics[angle=270,width=9.0cm]{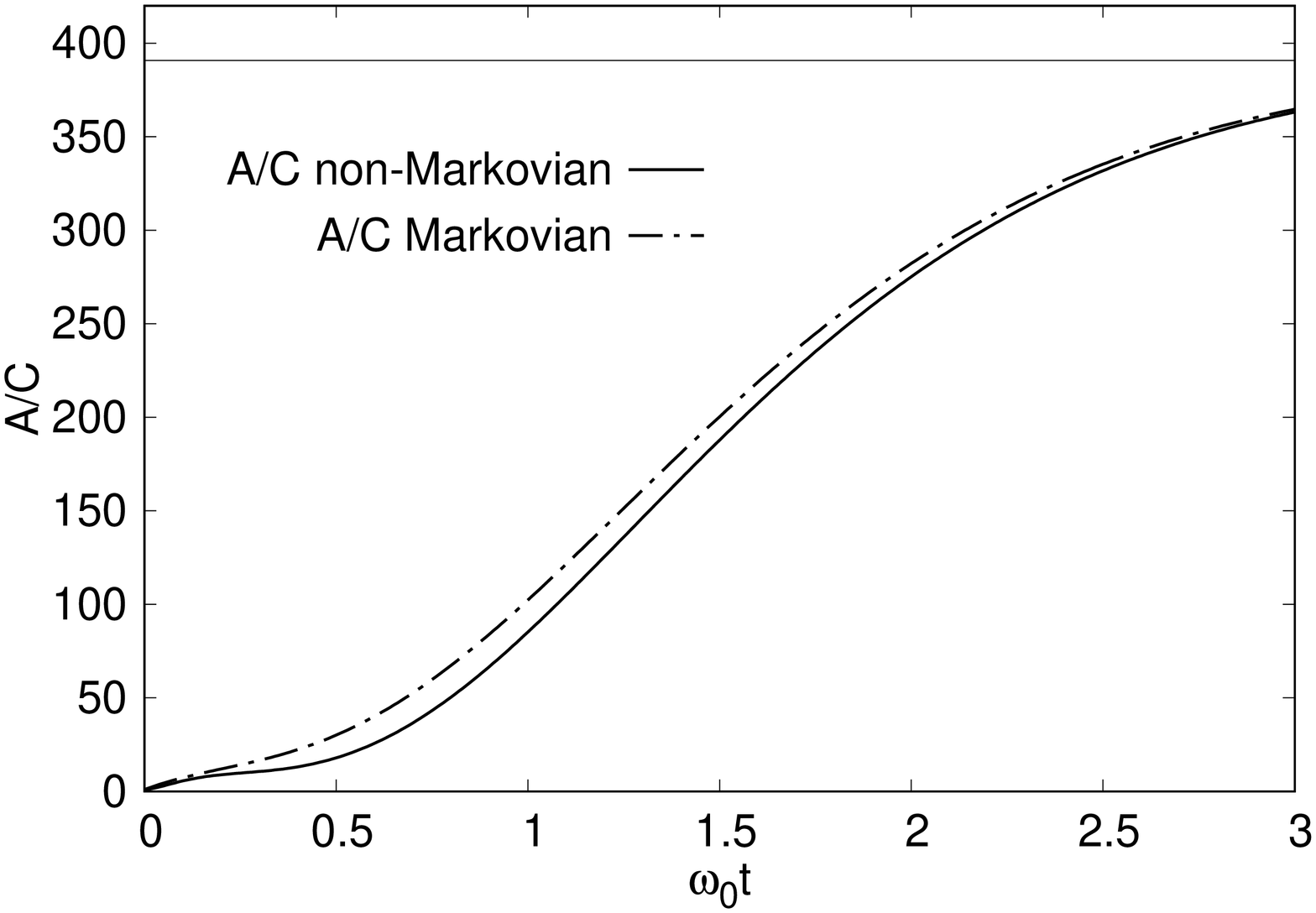}
			\caption{}
		\end{subfigure}
		\caption{The parameters used here are $\gamma=\omega_0$, $\Omega=20\omega_0$, and $k_BT=10\hbar\omega_0$. The initial conditions are $w=1$, $(c_1(0),c_2(0),c_3(0))=(d_0^2/4,0,1/\left(4 d_0^2\right))$. Left panel: 
			$A d_0^2$ and $C d_0^2$ as a function of $\omega_0t$, where $d_0$ is the width of the quantum harmonic oscillator's ground state. The main figure shows the non-Markovian time evolution and the inset shows 
			the Markovian time evolution. Right panel: $A/C$ as a function of $\omega_0t$. The solid and dash-dotted lines show this ratio for the non-Markovian and  the Markovian case, respectively. 
			The horizontal thin lines indicate the asymptotic values in both panels.} \label{case1}
	\end{figure*}
	
	\begin{figure*}[ht!]
		\begin{subfigure}{.48\textwidth}
			\includegraphics[angle=270,width=9cm]{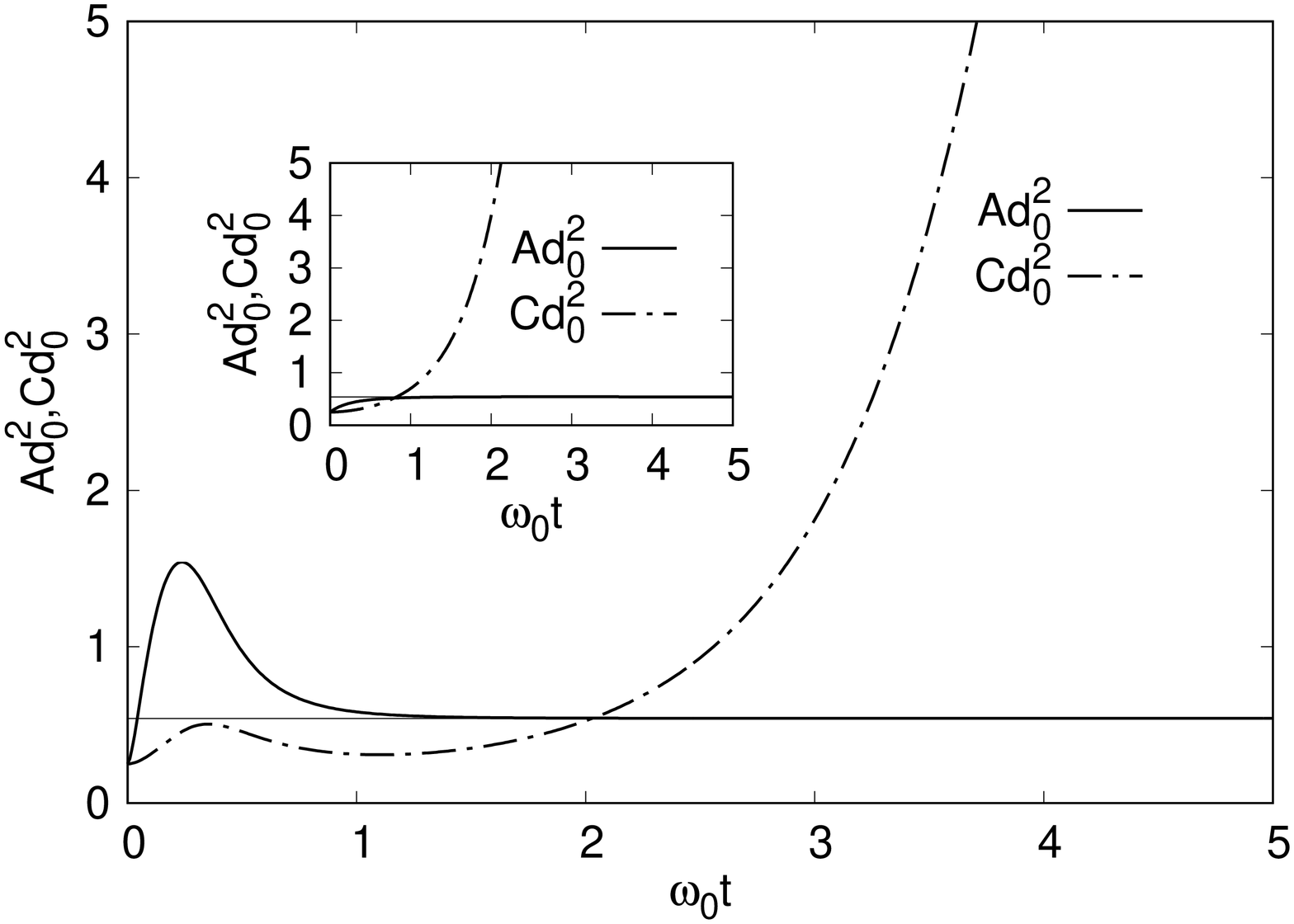}
			\caption{}
		\end{subfigure}\hspace{1em}
		\begin{subfigure}{.48\textwidth}
			\includegraphics[angle=270,width=9cm]{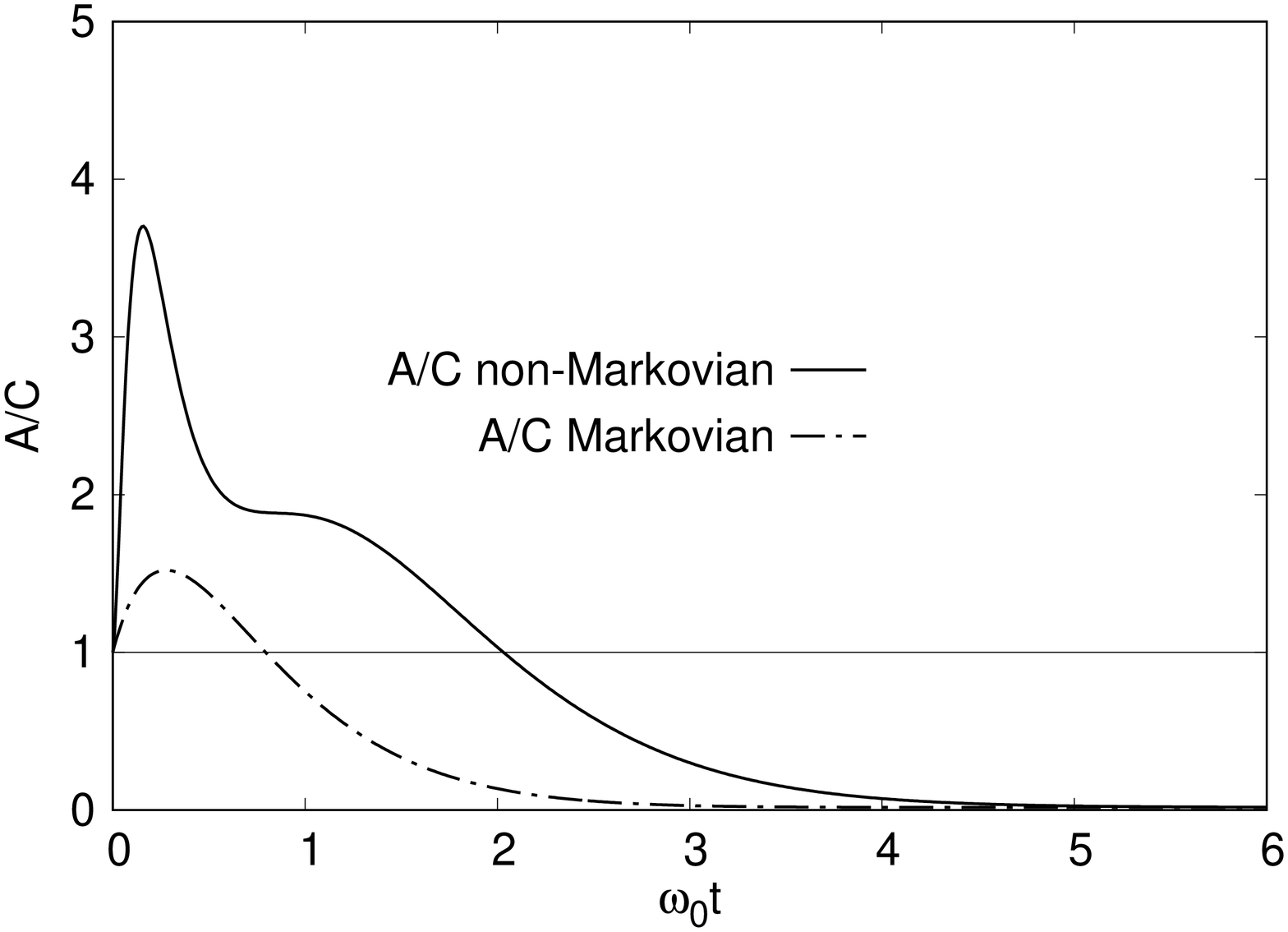}
			\caption{}
		\end{subfigure}
		\caption{The same as for Fig.~\ref{case1}. The parameters used here are $\gamma=\omega_0$, $\Omega=20\omega_0$, and $k_BT=\hbar \omega_0$. The initial conditions are $w=1$, $(c_1(0),c_2(0),c_3(0))=(d_0^2/4,0,1/\left(4 d_0^2\right))$.  
			Positivity violations occur for $\omega_0 t>0.79$ (Markovian case) and for $\omega_0 t> 2.03$ (non-Markovian case).} \label{case2}
	\end{figure*}
	
	\begin{figure*}[ht!]
		\begin{subfigure}{.48\textwidth}
			\includegraphics[angle=270,width=9cm]{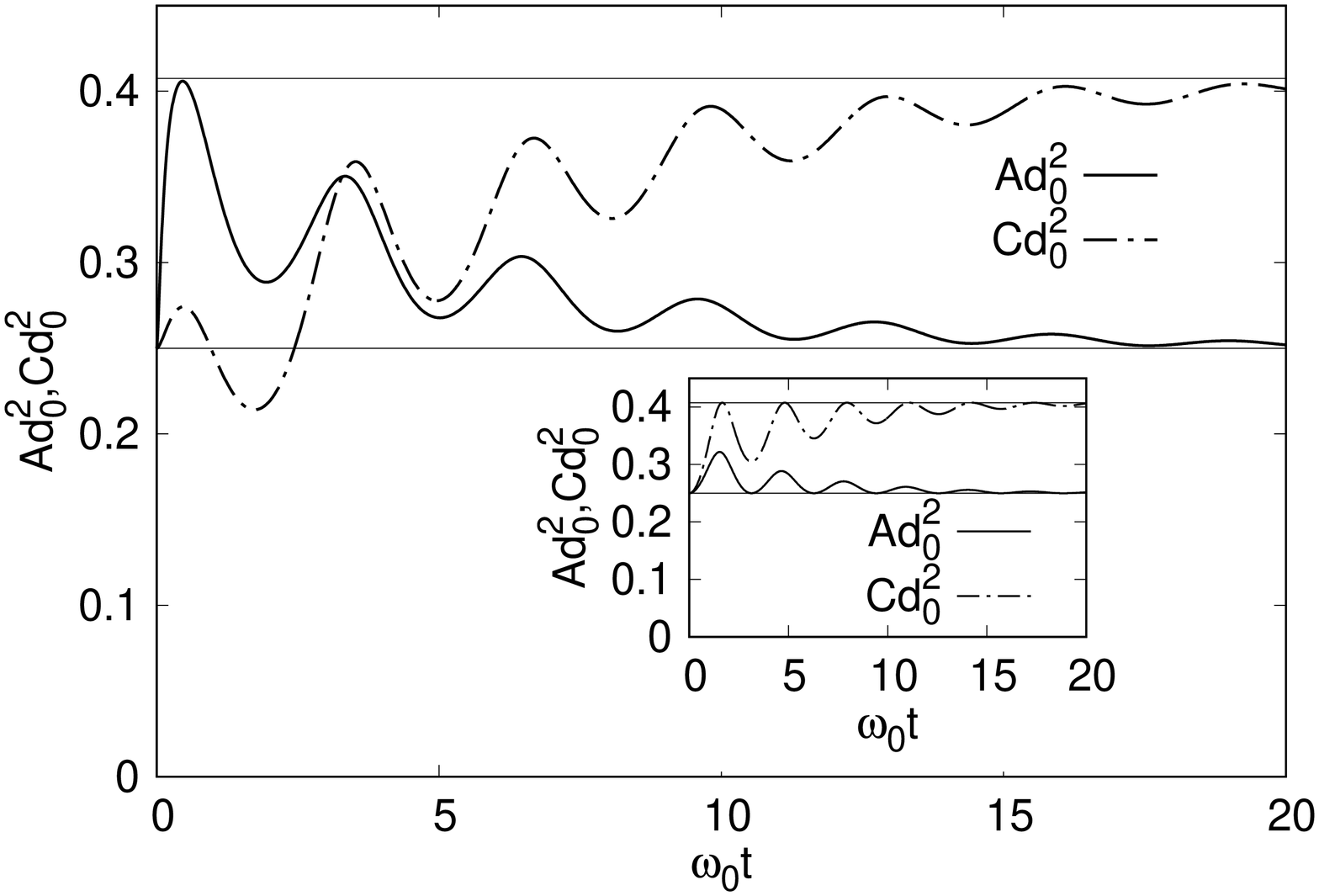}
			\caption{}
		\end{subfigure}\hspace{1em}
		\begin{subfigure}{.48\textwidth}
			\includegraphics[angle=270,width=9cm]{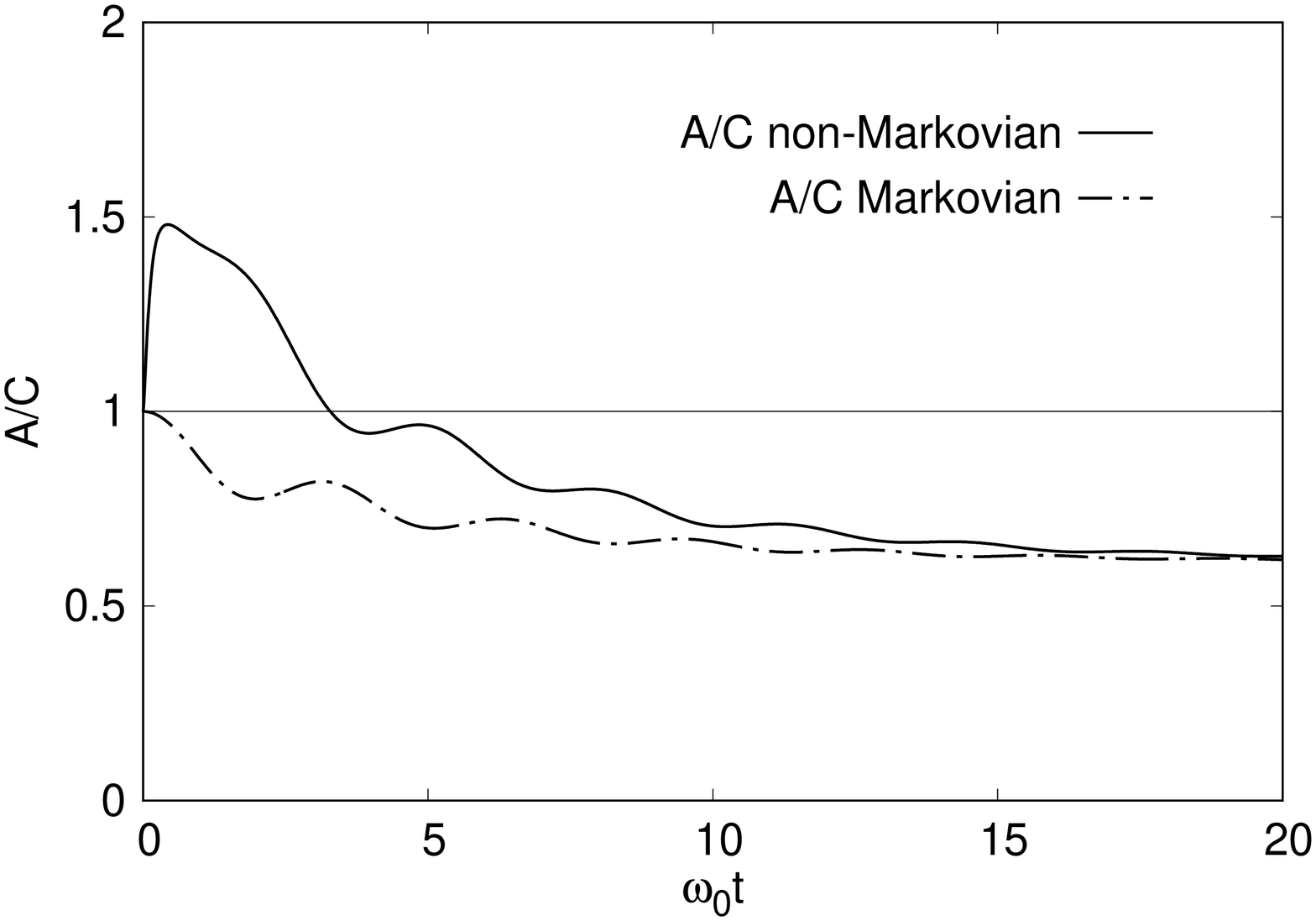}
			\caption{}
		\end{subfigure}
		\caption{The same as for Fig.~\ref{case1}. The parameters used here are $\gamma=0.1\omega_0$, $\Omega=20\omega_0$, and $k_BT=0.01\hbar \omega_0$. The initial conditions are $w=1$, $(c_1(0),c_2(0),c_3(0))=(d_0^2/4,0,1/\left(4 d_0^2\right))$.
			Positivity violations occur for $\omega_0 t>0$ (Markovian case) and for $\omega_0 t> 3.28$ (non-Markovian case).
		} \label{case3}
	\end{figure*}
	
	A different approximation is possible for $\gamma_\mathrm{crit}$ at very low temperature.
	Keeping the leading-order terms in Eq.~\eqref{eq:gamma_crit_line} for $k_B T \ll \hbar\omega_0$ and $k_B T \ll \hbar\Omega$, one gets the limiting behavior
	\begin{eqnarray}
	\gamma_\mathrm{crit} \cong 4\frac{\frac{\Omega^2+\omega_0^2}{\Omega}\exp{\left\lbrace - \frac{\hbar \omega_0}{k_B T}\right\rbrace }}{2-\frac{4}{\pi}\frac{\Omega}{\omega_0}\ln\left( \frac{\omega_0}{\Omega}\right) }
	\equiv C \cdot e^{- \frac{a}{T}},
	\label{eq:gamma_for_small_T}
	\end{eqnarray}
	where $C$ and $a$ are constants. Clearly, this function is non analytical in $T$, and approaches the 
	origin in Fig.~\ref{density} with infinite slope. Inverting \eqref{eq:gamma_for_small_T} one has on the critical line 
	\begin{equation}
	T \cong \frac{a}{\ln(\frac{C}{\gamma_\mathrm{crit}})},
	\end{equation}
	for small $\gamma_\mathrm{crit}$.
	
	\begin{figure*}[ht!]
		\begin{subfigure}{.48\textwidth}
			\includegraphics[angle=270,width=9cm]{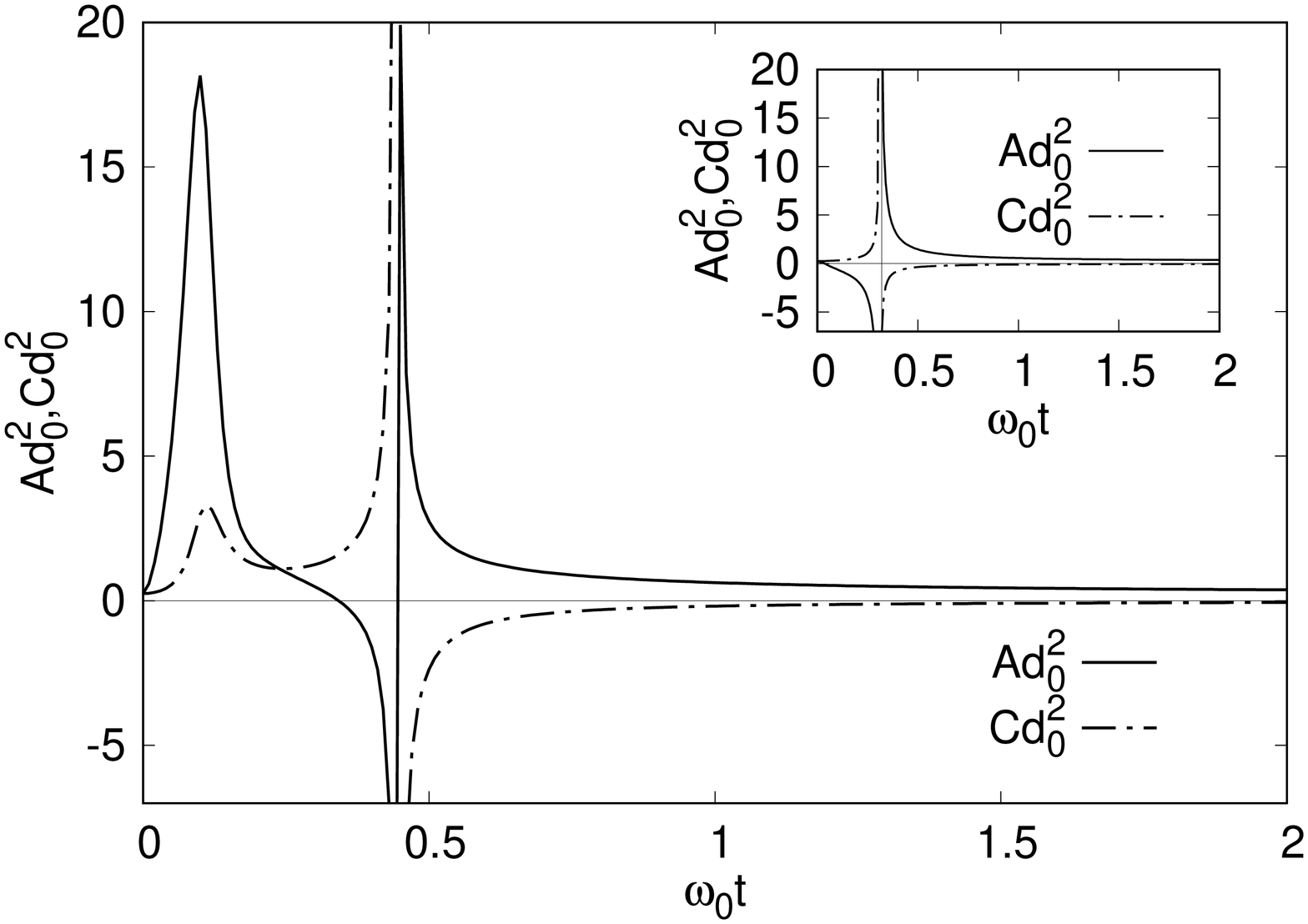}
			\caption{}
		\end{subfigure}\hspace{1em}
		\begin{subfigure}{.48\textwidth}
			\includegraphics[angle=270,width=9cm]{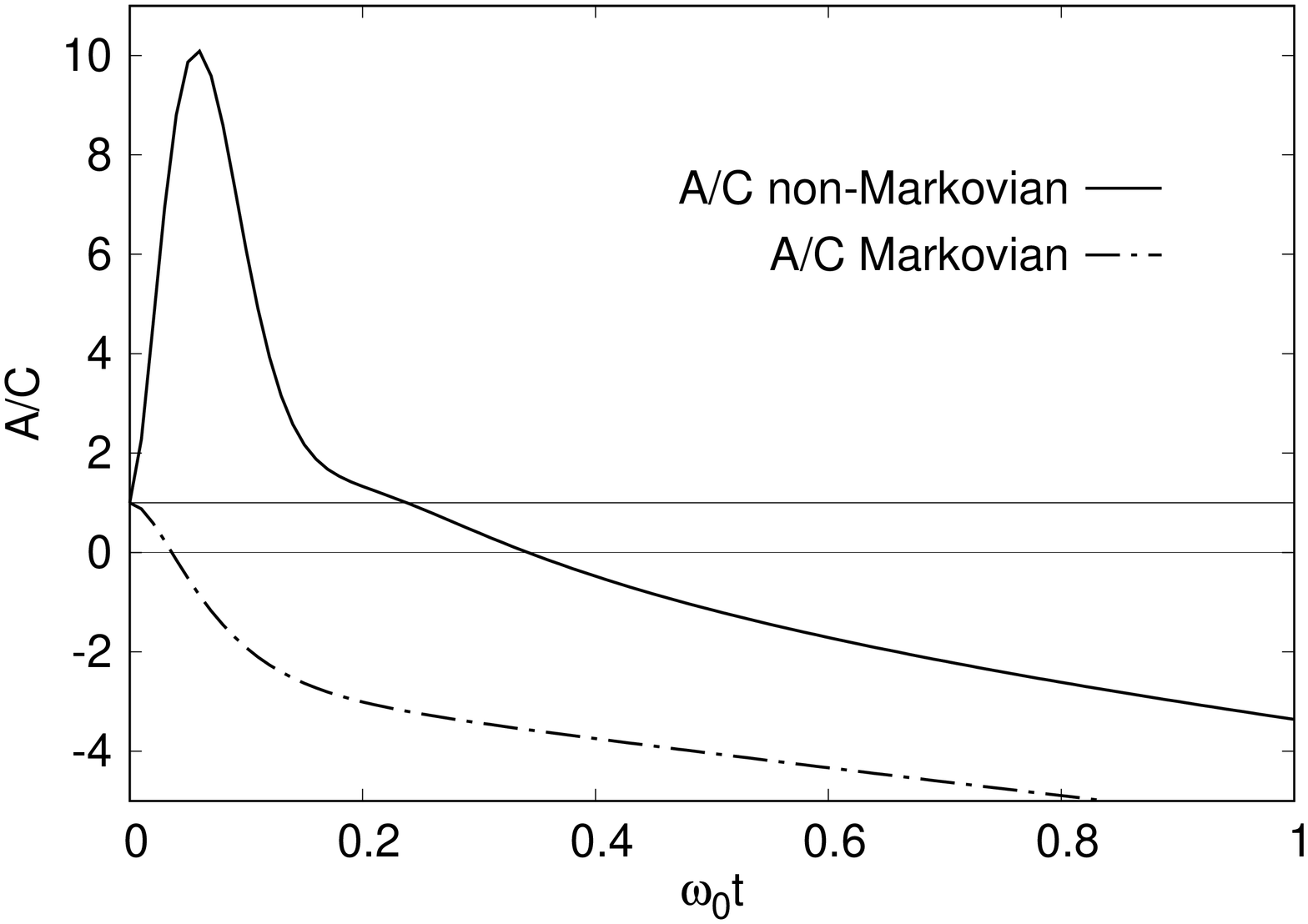}
			\caption{}
		\end{subfigure}
		\caption{The same as for Fig.~\ref{case1}. The parameters used here are $\gamma=10\omega_0$, $\Omega=20\omega_0$, and $k_BT=0.01\hbar \omega_0$. The initial conditions are $w=1$, $(c_1(0),c_2(0),c_3(0))=(d_0^2/4,0,1/\left(4 d_0^2\right))$. 
			Positivity violations occur for $\omega_0 t>0$ (Markovian case) and for $\omega_0 t> 0.23$ (non-Markovian case). For the non-Markovian case $A$ changes sign at $\omega_0t \approx 0.36$. $A$ and $C$ 
			diverge at $\omega_0t \approx 0.44$.} \label{case4}
	\end{figure*}
	
	As we indicated earlier, one can experience positivity violations during the 
	time evolution. In the following, we show a few time evolutions which might be interesting for the reader. In the numerics, we limit ourselves to Gaussian density operators, which means that 
	we have to follow only the time evolution of $\mathbf{c}(t)$, from which we extract  $A$ and $C$ via \eqref{eq:A_and_C_for_cs} and check the validity of \eqref{good_spectrum} numerically.
	
	\begin{figure*}[ht!]
		\begin{subfigure}{.48\textwidth}
			\includegraphics[angle=270,width=9cm]{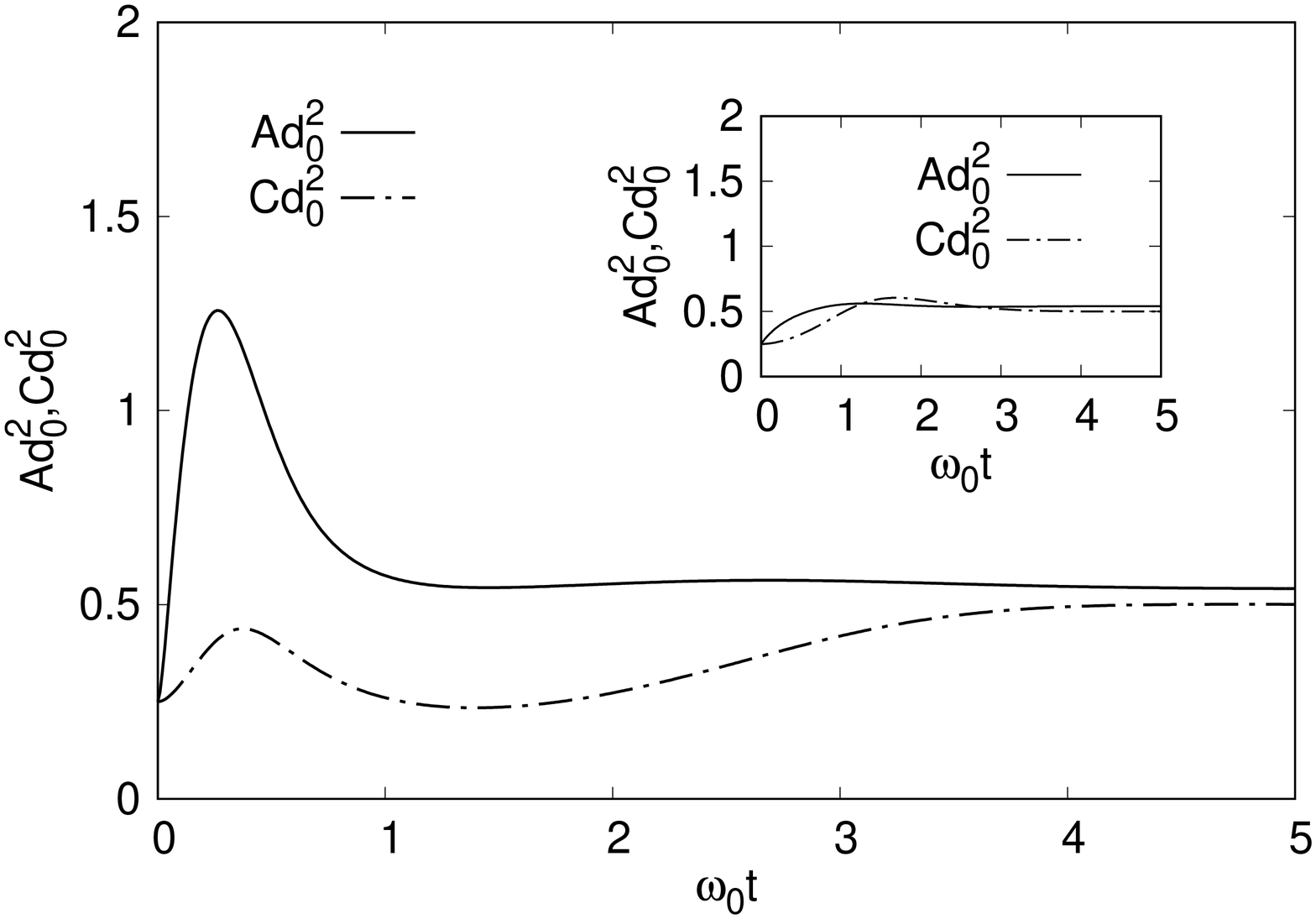}
			\caption{}
		\end{subfigure}\hspace{1em}
		\begin{subfigure}{.48\textwidth}
			\includegraphics[angle=270,width=9cm]{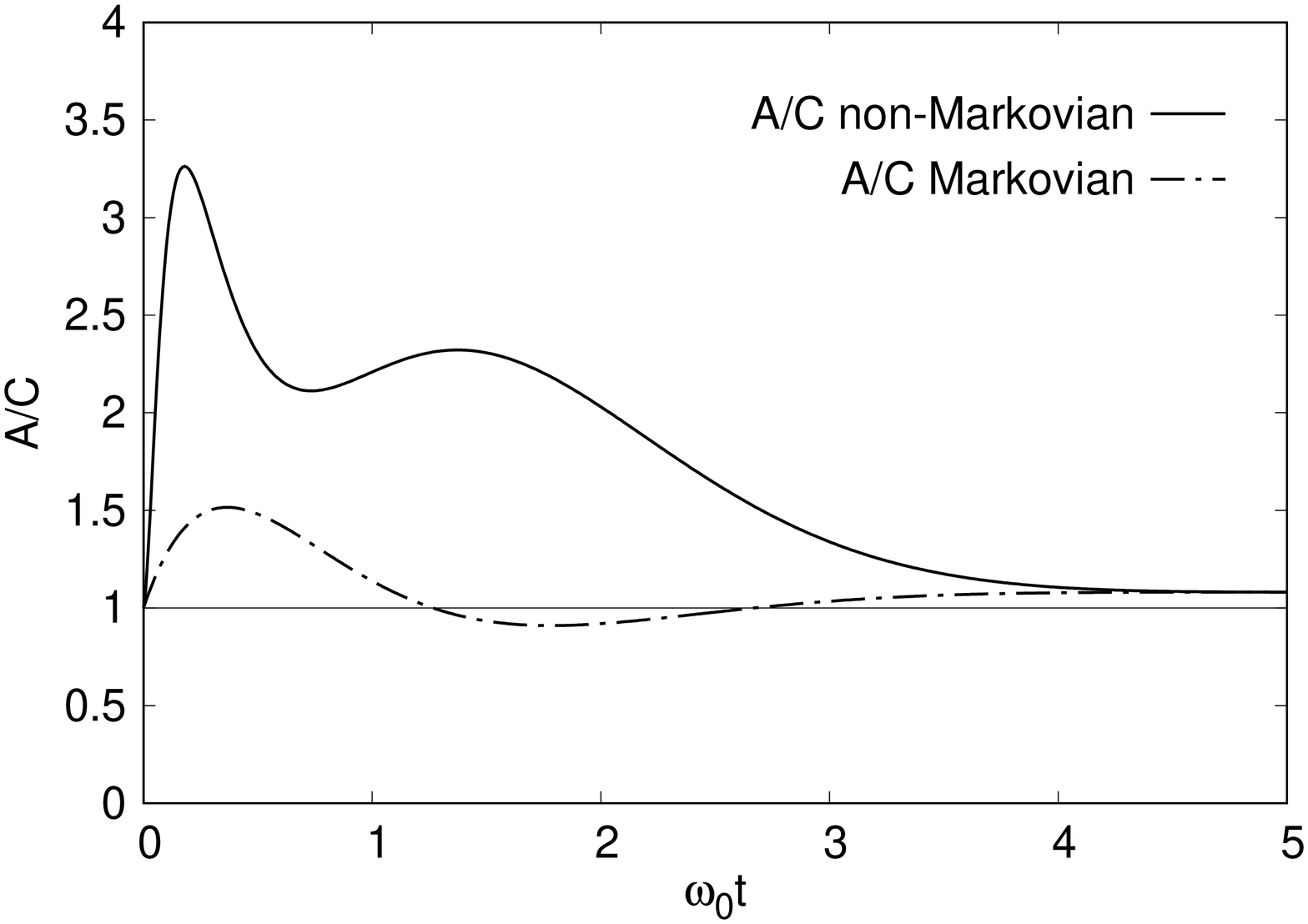}
			\caption{}
		\end{subfigure}
		\caption{The same as for Fig.~\ref{case1}. T he parameters used here are $\gamma=0.755\omega_0$, $\Omega=20\omega_0$, and $k_BT=\hbar \omega_0$. The  initial conditions are: $w=1$, $(c_1(0),c_2(0),c_3(0))=
			(d_0^2/4,0,1/\left(4 d_0^2\right))$. 
			The Markovian behavior is unphysical for $1.26 < \omega_0 t < 2.68$.} \label{case5}
	\end{figure*}
	
	\begin{figure}[ht!]
		\includegraphics[angle=270,width=9cm]{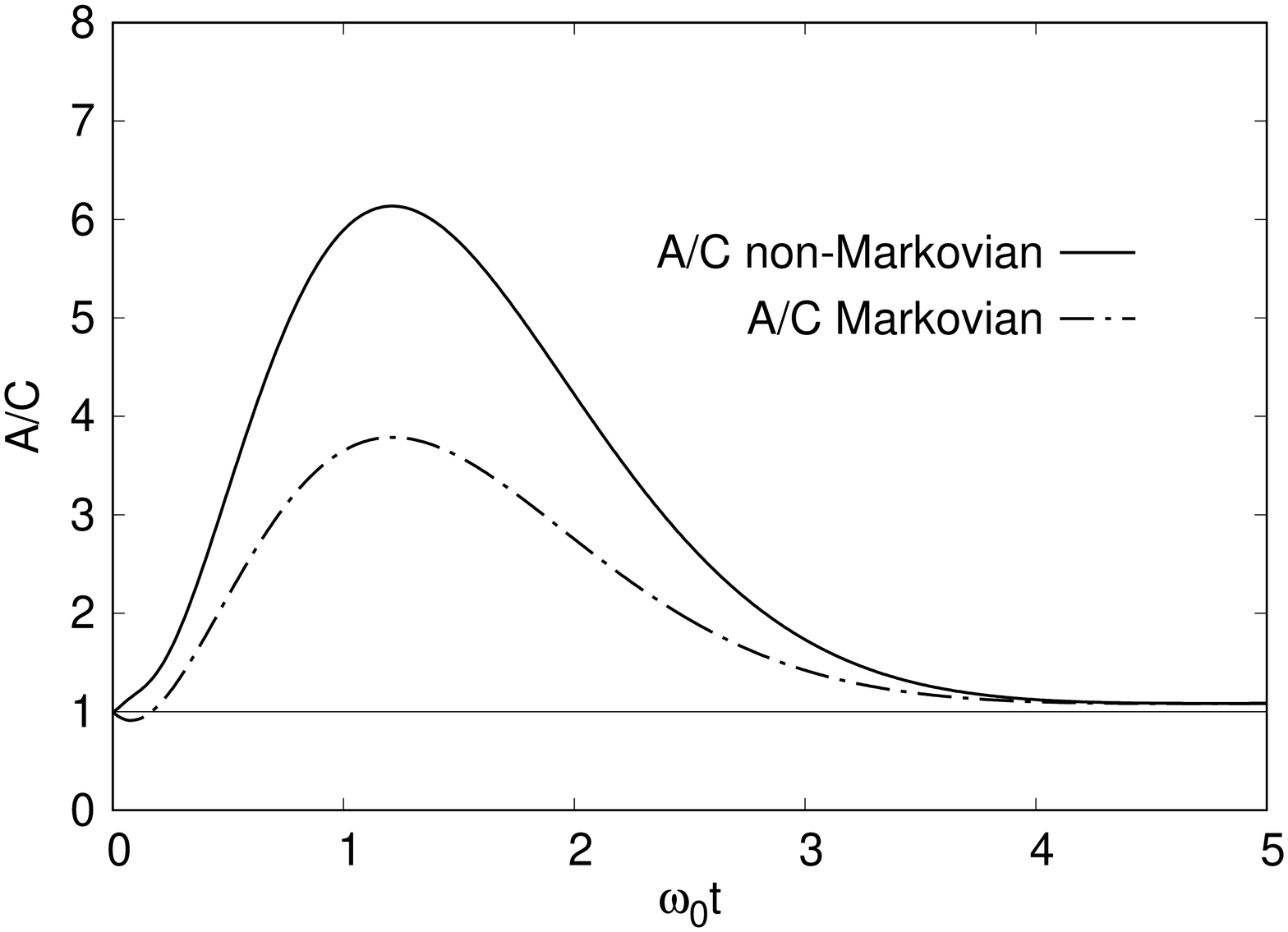}
		\caption{The same as for Fig.~\ref{case1}(b). the parameters used here are $\gamma=0.755\omega_0$, $\Omega=20\omega_0$, and $k_BT=\hbar \omega_0$. The initial conditions are $w=1/\sqrt{10}$, $(c_1(0),c_2(0),c_3(0))=
			(d_0^2/(40),0,10/\left(4 d_0^2\right))$. 
			The Markovian behavior is unphysical for $0 < \omega_0 t < 0.17$.} \label{case20}
	\end{figure}
	
	\begin{figure*}[ht!]
		\begin{subfigure}{.48\textwidth}
			\includegraphics[angle=270,width=9cm]{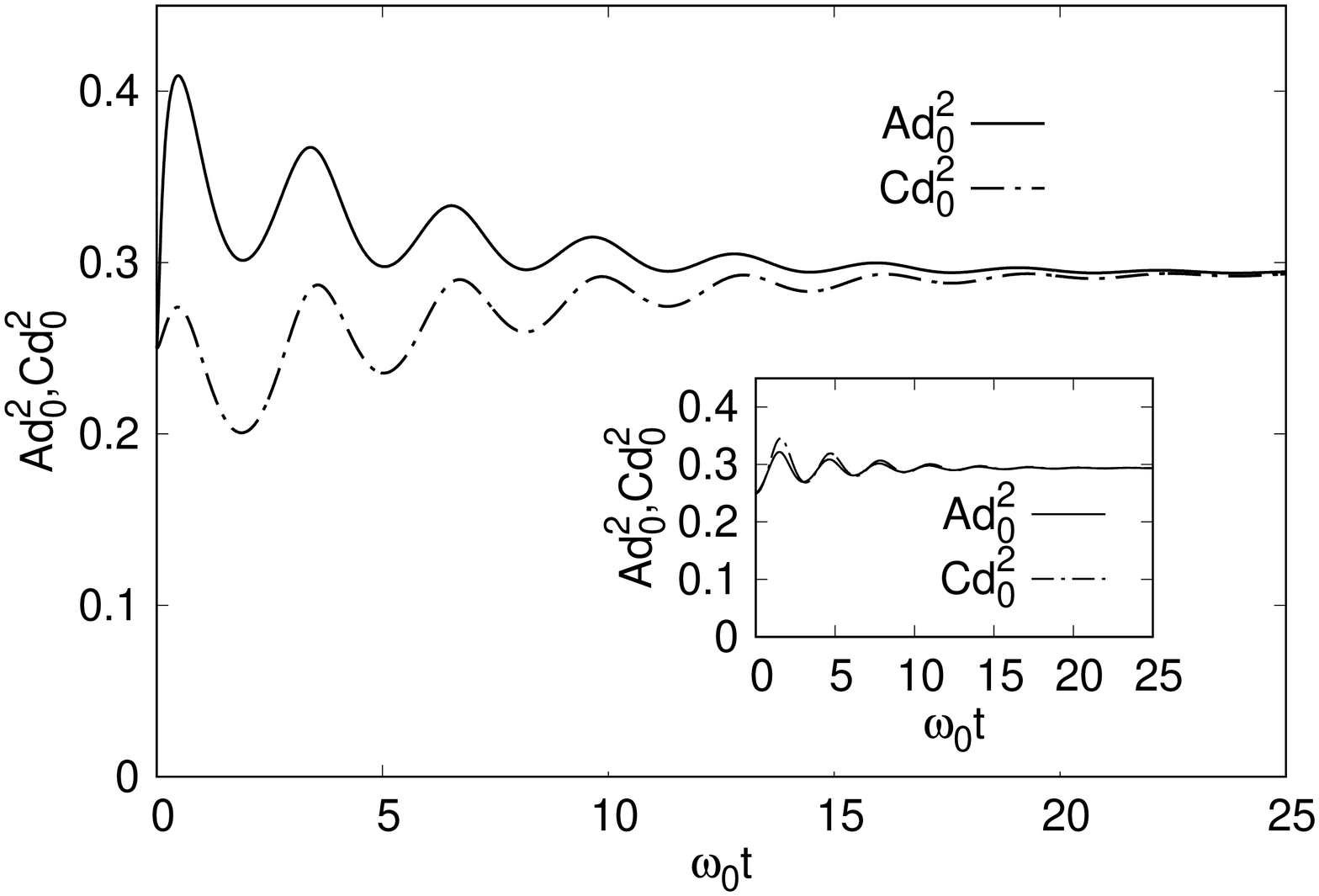}
			\caption{}
		\end{subfigure}\hspace{1em}
		\begin{subfigure}{.48\textwidth}
			\includegraphics[angle=270,width=9cm]{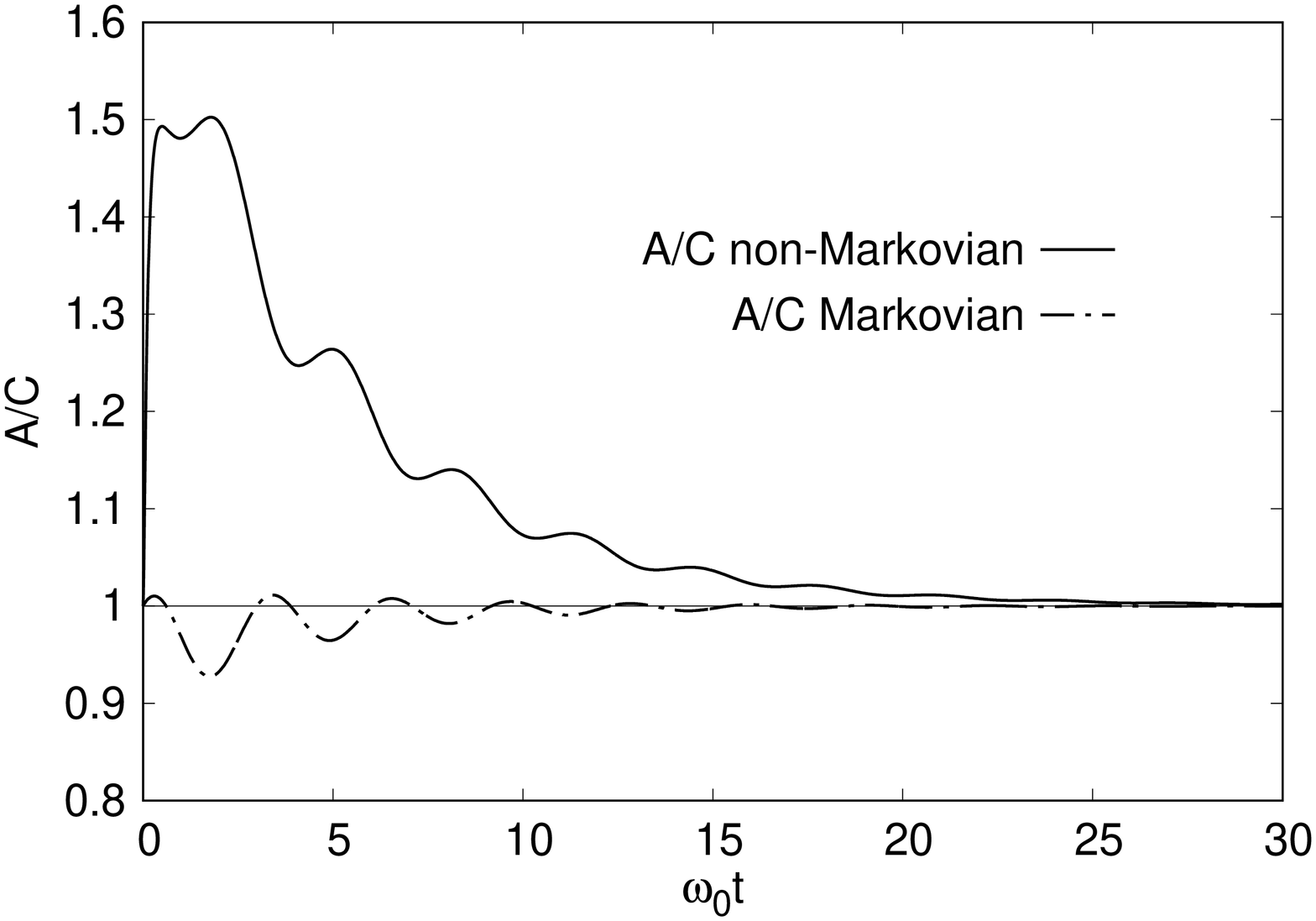}
			\caption{}
		\end{subfigure}
		\caption{The same as for Fig.~\ref{case1}. The parameters used here are $\gamma=0.1\omega_0$, $\Omega=20\omega_0$, $k_BT=0.397055\hbar \omega_0$. The initial conditions are $w=1$, 
			$(c_1(0),c_2(0),c_3(0))=(d_0^2/4,0,1/\left(4 d_0^2\right))$. 
			Several positivity violations are for the Markovian case.} \label{case11}
	\end{figure*}
	
	\begin{figure}[ht!]
		\includegraphics[angle=270,width=9cm]{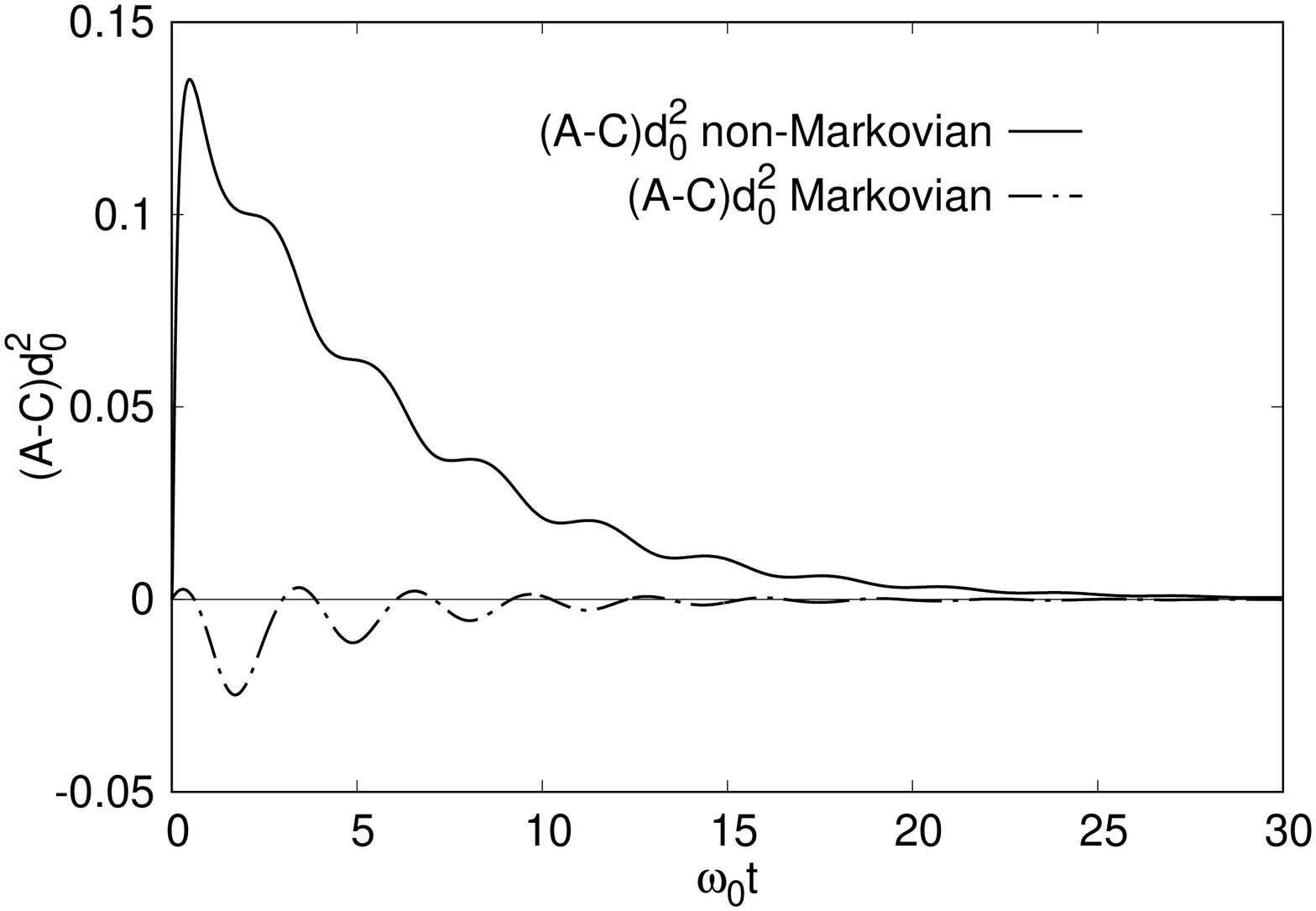}
		\caption{$(A-C)d_0^2$ as a function of $\omega_0t$. The parameters used here are $\gamma=0.1\omega_0$, $\Omega=20\omega_0$, $k_BT=0.397055\hbar \omega_0$. The initial conditions are $w=1$, $(c_1(0),c_2(0),c_3(0))=
			(d_0^2/4,0,1/\left(4 d_0^2\right))$. Solid line: non-Markovian case; dash-dotted line: Markovian case. The horizontal line is drawn at zero. One can observe several positivity violations for the Markovian case.} 
		\label{case11_minus}
	\end{figure}

	\begin{figure}[ht!]
		\includegraphics[width=8cm]{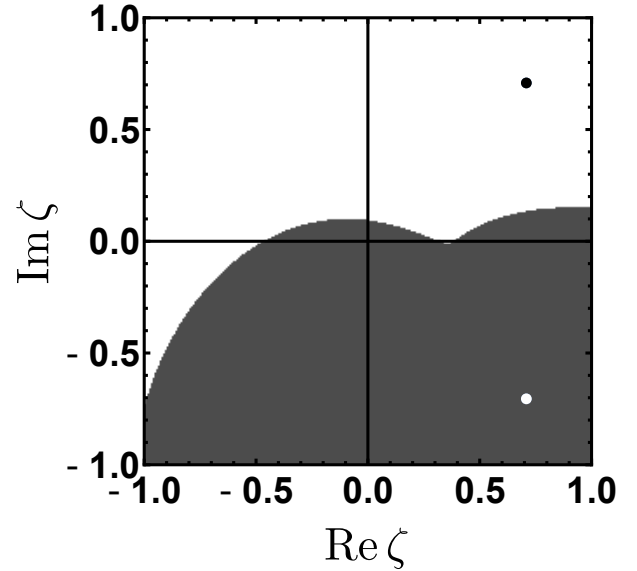}
		\caption{Positivity violations in the Markovian runs for different squeezed initial conditions characterized by the complex $\zeta$. White region: no positivity violations, dark region: positivity violations. 
			The parameters used here are  $\gamma=0.755\omega_0$, $\Omega=20\omega_0$, $k_BT=\hbar \omega_0$, $w=1$. Time evolutions for points at $|\zeta|=1$, $\phi=\pm \pi/4$ will be shown on Fig.\ref{case25}.}
		\label{fig:zeta_phi}
	\end{figure}

	\begin{figure*}[ht!]
		\begin{subfigure}{.48\textwidth}
			\includegraphics[angle=270,width=9cm]{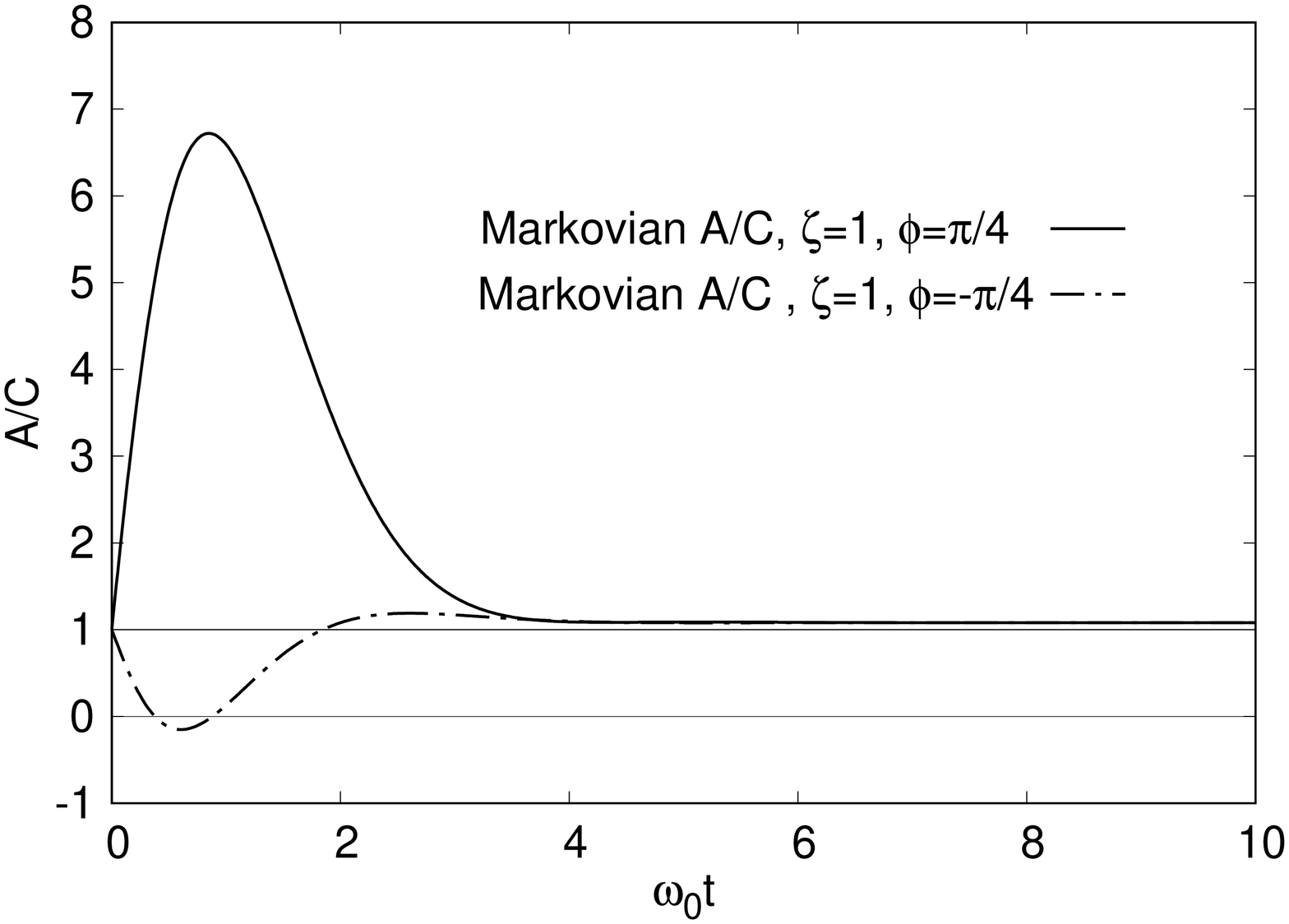}
			\caption{}
		\end{subfigure}\hspace{1em}
		\begin{subfigure}{.48\textwidth}
			\includegraphics[angle=270,width=9cm]{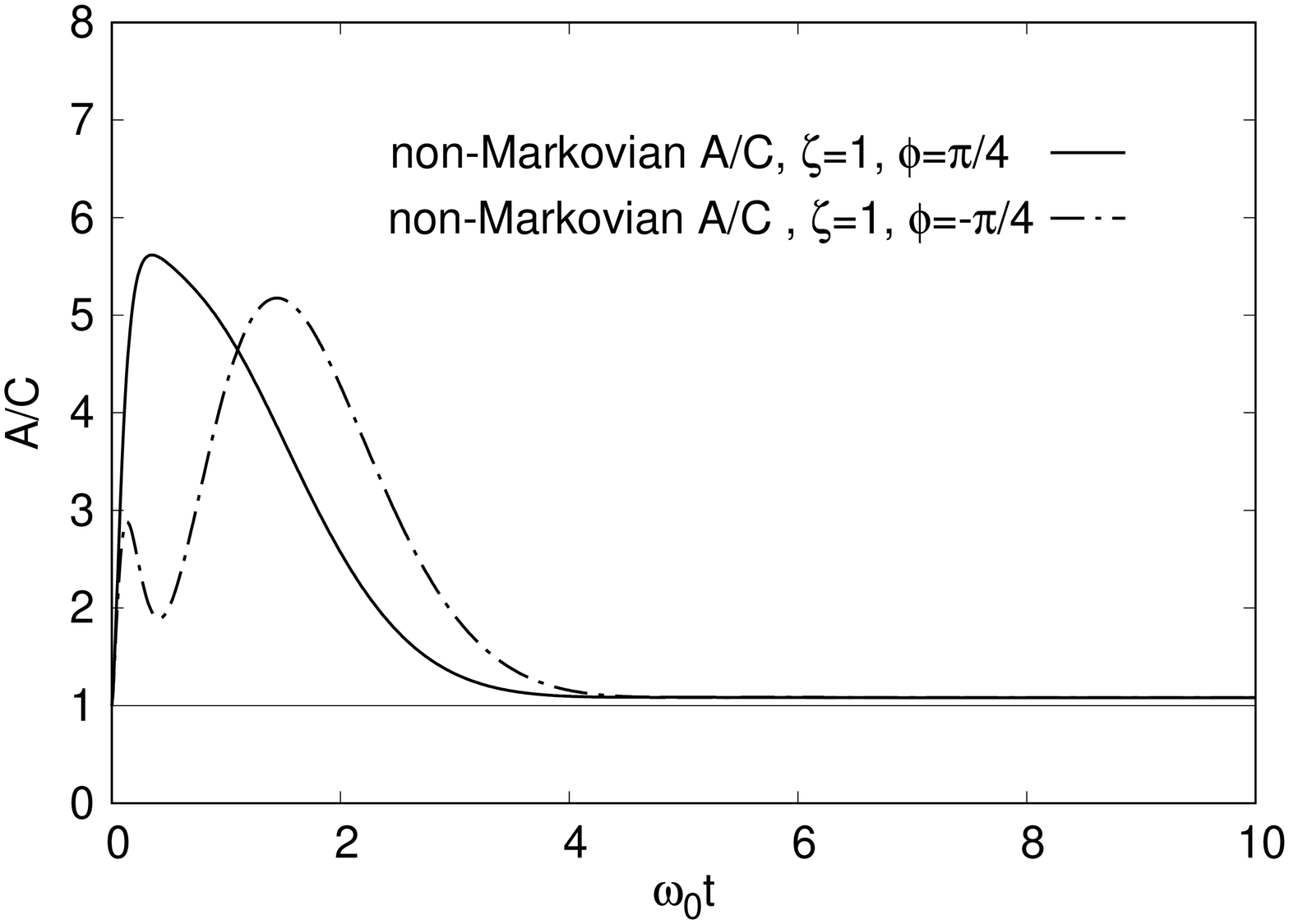}
			\caption{}
		\end{subfigure}
		\caption{$A(t)/C(t)$ for selected squeezed initial states. Left Panel: Markovian time evolutions, right panel: non-Markovian time evolutions.
			The parameters used here are $\gamma=0.755\omega_0$, $\Omega=20\omega_0$, and $k_BT=\hbar \omega_0$. The initial conditions are $w=1$, $\zeta=1$, $\phi=\pm \pi/4$, $(c_1(0),c_2(0),c_3(0))=(0.299405\,d_0^2,\pm 1.282289,1.581693/d_0^2)$. 
			Positivity violation is in the interval $0<\omega_0 t<1.85$ for the Markovian case, with $\phi=-\pi/4$. No violation is on the right panel.} \label{case25} 
	\end{figure*}

	\begin{figure}[ht!]
		\includegraphics[angle=0,width=9cm]{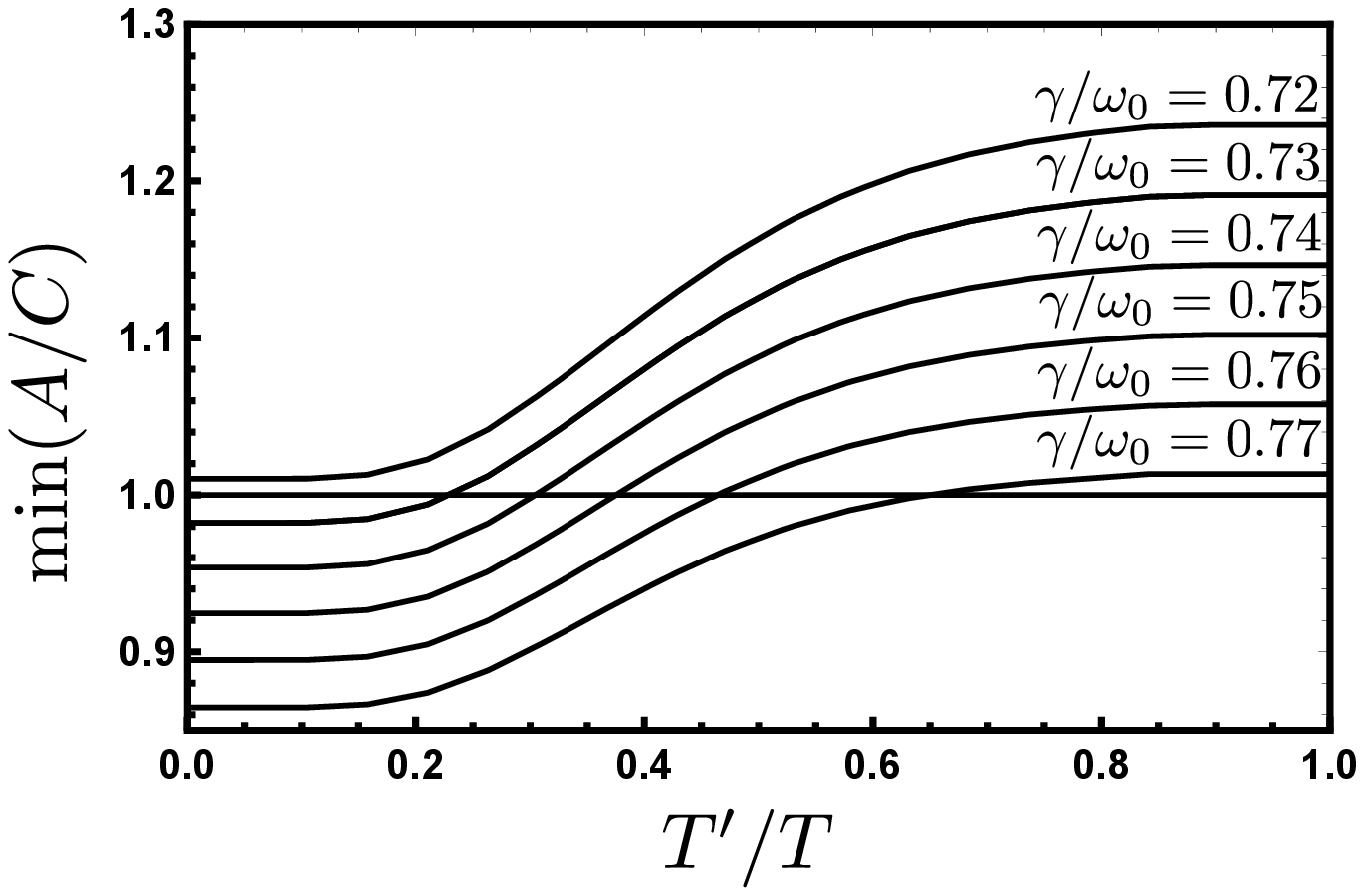}
		\caption{Behavior of $\min_{t\ge 0} A(t)/C(t)$ for thermal initial states \eqref{thermalstate} of temperature $T'$ for the Markovian time evolution. Various values of $\gamma$ are used and we set
			$\Omega=20\omega_0$ and $k_BT=\hbar\omega_0$.} 
		\label{thermal_min}
	\end{figure}
	
	\section{Numerical results}
	\label{IV}
	
	In the previous section, we have discussed the validity of the stationary solution, which gives a constraint on the parameters of the master equation.
	We consider three different types of initial conditions of \eqref{eq:main_diffeq}, namely, coherent, squeezed, and thermal states. 
	For the sake of completeness we hereby reformulate these well-known  initial states to our representation. 
	
	{\it Coherent state}. This state is defined through the complex parameter $\alpha$,
	\begin{equation}
	\label{coherent}
	\ket{\alpha}=\sum_{n=0}^\infty 
	e^{-\frac{|\alpha|^2}{2}}
	\frac{\alpha^n}{\sqrt{n!}}
	\ket n,
	\quad\alpha=|\alpha|e^{i\phi},
	\end{equation}
	where $\ket{n}$ ($n\in {\mathbb N}_0$) are the number states and $\phi$ is the complex phase of $\alpha$. The Wigner function of this coherent state reads
	\begin{equation}
	W(x,p)=\frac{1}{\pi \hbar} e^{-\left(x/d-\sqrt{2} \mathrm{Re}(\alpha)  \right)^2-\left(pd/\hbar-\sqrt{2} \mathrm{Im}(\alpha)  \right)^2} \nonumber
	\end{equation}
	where $d$ is a length and can be taken as
	\[
	d=w\sqrt{\frac{\hbar}{m\omega_0}}\equiv w d_0.
	\]
	$w$ is a dimensionless positive number, and $d_0$ is the width of the quantum harmonic oscillator's ground state. Due to the relation
	\begin{equation}
	W(x,p)=\left( \frac{1}{2 \pi}\right) ^2\int^{\infty}_{-\infty} dk \int^{\infty}_{-\infty} d \Delta e^{-i(kx+\Delta p/\hbar)}  \rho(k, \Delta) \label{transform2}
	\end{equation}
	we obtain
	\begin{equation}
	\mathbf{c}_\mathrm{coh}(0)=\left(\frac{d^2}{4},0,\frac{1}{4 d^2}\right). 
	\label{eq:coherent_init_cond}
	\end{equation}
	
	{\it Squeezed state}.  In this case the state is characterized by two complex parameters $\alpha$ and $\zeta=|\zeta|e^{i\phi}$. Introducing the creation $a^\dagger$ and annihilation $a$ operators of 
	the quantum harmonic oscillator a squeezed state is given by
	\begin{equation}
	\ket{\alpha,\zeta}=\hat{D}(\alpha)\hat{S}(\zeta)\ket{0}
	\label{squeezed}
	\end{equation}
	where $\hat{D}(\alpha)=\exp\bigl(\alpha\hat{a}^\dagger-\alpha^\ast\hat{a}\bigr)$ is the displacement and $\hat{S}(\zeta)=\exp\bigl[\tfrac{1}{2}
	\bigl(\zeta^\ast{}a^2-\zeta{}a^{\dagger 2}\bigr)\bigr]$ is the squeezing operator. After a lengthy but standard calculation, the Wigner function yields
	\begin{eqnarray}
	W(x,p)&=&\frac{1}{\pi \hbar} e^{-\left(x/d-\sqrt{2} \mathrm{Re}(\alpha)  \right)^2 t_1-\left(pd/\hbar-\sqrt{2} \mathrm{Im}(\alpha)  \right)^2 t_2} \nonumber \\
	&\times& e^{\left(x/d-\sqrt{2} \mathrm{Re}(\alpha)  \right)\left(pd/\hbar-\sqrt{2} \mathrm{Im}(\alpha)  \right)t_3 } \nonumber
	\end{eqnarray}
	where 
	\begin{eqnarray}
	t_1&=&\frac{e^{2|\zeta|}}{2}(1+\cos \phi)+\frac{e^{-2|\zeta|}}{2}(1-\cos \phi), \nonumber \\
	t_2&=&\frac{e^{2|\zeta|}}{2}(1-\cos \phi)+\frac{e^{-2|\zeta|}}{2}(1+\cos \phi), \nonumber \\
	t_3&=&\left( e^{2|\zeta|}-e^{-2|\zeta|}\right) \sin \phi. \label{eq:squeezed_params}
	\end{eqnarray}
	Finally, with the help of \eqref{transform2}, we get
	\begin{equation}
	\mathbf{c}_\mathrm{sq}(0)=\left(\frac{d^2}{4}t_2,\frac{1}{4}t_3,\frac{1}{4 d^2}t_1\right). 
	\end{equation}
	
	{\it Thermal state}. This is a Gibbs state characterized by the thermal equilibrium temperature $T'$, which in the number state representation reads
	\begin{equation}
	\hat{\rho}=\sum_n \frac{ n^n_{\text{th}}}{(1+ n_{\text{th}})^n} \ket{n}\bra{n}
	\label{thermal}
	\end{equation}
	with the mean excitation number
	\begin{equation}
	n_{\text{th}} = \biggl[\exp \biggl( \frac{\hbar\omega_0}{k_\text{B}T'} \biggr)-1 \biggr]^{-1}. \nonumber
	\end{equation}
	We have for the Wigner function
	\begin{equation}
	W(x,p)=\frac{1}{\pi \hbar} e^{-\frac{x^2}{d^2 (2 n_{\text{th}}+1)}-\frac{p^2 d^2}{\hbar^2 (2 n_{\text{th}}+1)} } \nonumber
	\end{equation}
	which yields
	\begin{equation}\label{thermalstate}
	\mathbf{c}_\mathrm{th}(0)=\coth\left(\frac{\hbar\omega_0}{2k_B T'} \right)\left(\frac{d^2}{4},0,\frac{1}{4 d^2}\right).
	\end{equation}
	Note that the coherent state with $w=1$ corresponds to the ground state of the quantum harmonic oscillator and is contained as trivial special cases of the thermal and squeezed states.
	
	In all subsequent numerical cases we will compare the time evolution of \eqref{eq:main_diffeq} with its Markovian version which is obtained by replacing all time-dependent 
	coefficient functions with their respective limits as $t \to \infty$ i.e., 
	\begin{eqnarray}
	&&\omega_p(t) \rightarrow \omega^{(M)}_p, \quad \lambda(t) \rightarrow \lambda^{(M)},\nonumber \\
	&&D_{px}(t)\rightarrow D^{(M)}_{px},\quad D_{pp}(t)\rightarrow D^{(M)}_{pp}. \nonumber
	\end{eqnarray}
	
	The result of a typical, physically valid time evolution can be seen in Fig.~\ref{case1}. Here the parameters are chosen so that the density operator is physical for any time, i.e.,
	$A$ and $C$ are positive and $A \geq C$. One can observe very similar behavior if one starts from a squeezed or a thermal state, except  $A/C$ starts from a number bigger than 1 for a thermal state. In the 
	figures, we use dimensionless units, $A$ and $C$ are multiplied with $d_0^2$, where $d_0$ is the width of the quantum harmonic oscillator's ground state.
	
	In Fig.~\ref{case2}, parameters are chosen from region I. It promptly follows that the asymptotic behavior must be unphysical for both the Markovian and non-Markovian cases. 
	In Fig.~\ref{case2}(b), both curves are already below the horizontal line at $\omega_0 t \approx 2$; however, the duration of the physical behavior is longer for the non-Markovian case at the beginning.  
	The parameters  in Fig.~\ref{case3} are also from region I; however, the 
	comparison with the previous case shows that for smaller temperature $k_BT/\hbar \omega_0$, and damping factor $\gamma/\omega_0$ we can see a few oscillations. 
	The non-Markovian evolution is physical up to $\omega_0t \approx 3.28$ and later it becomes unphysical because $A/C$ becomes 
	smaller than one. The Markovian evolution promptly becomes unphysical at $t=0$ and remains for all times. We note that the parameters $\gamma/\omega_0$, $\Omega/\omega_0$, and $k_BT/\hbar \omega_0$ are chosen 
	to be the same as for the bottom subfigure of Fig.~(10.7) in the book by Breuer and Petruccione \cite{book1}.    
	
	In Fig.~\ref{case4}, we choose a bigger $\gamma$ than in Fig.~\ref{case3}. All of the other parameters and initial conditions are the same. The parameters still belong to region I. Here, something more drastic
	happens in both cases. First the ratio of $A/C$ goes below 1 (indicating positivity violation) and, at a later time, $A$ changes sign and  at an even further time, $A$ and, $C$ diverge, changing signs anew. 
	The Markovian evolution is still 
	unphysical for the whole time evolution, while non-Markovian evolution shows physical behavior until $A/C$ goes below one. If any of $A$ and $C$ become negative, the corresponding 
	Wigner function and $\textrm{Tr}\,\hat{\rho}$ do not exist. 
	
	In Fig.~\ref{case5}, we used the same parameters as in Fig.~\ref{case2}, except that $\gamma$ has been decreased in such a way that 
	the parameters are now in region II. 
	The non-Markovian evolution is physical for all time. The Markovian evolution gets unphysical but bounces back into the $A/C \geqslant 1$ region and remains physical at later times.

	For Fig.~\ref{case5}, the initial condition is a coherent state with $w=1$ in Eq.~(\ref{eq:coherent_init_cond}).
	It is interesting to note that if we vary $w$, for example to $w=1/\sqrt{10}$ the initial behavior of the Markovian run is completely different (see Fig.~\ref{case20}): 
	the positivity is promptly violated at $t=0^+$ and, at a later time, the system returns back to a physically allowed state. The non-Markovian time evolution remains physical for all the time even for 
	this initial condition. 
	
	An interesting regime is when $\gamma/\omega_0$ and $k_BT/\hbar \omega_0$ are small. Here we expect a few damped oscillations. In Figs.~\ref{case11} and \ref{case11_minus}, our parameters are close to the 
	critical line, but are still in region II.  The non-Markovian time evolution is already physical at any time. 
	However, the Markovian run shows several time intervals where the curve of $A/C$ attains values smaller than one. The same can also be monitored  in the quantity $A-C$  (see Fig.~\ref{case11_minus}).
	
	Let us discuss a few facts about squeezed initial states. Choosing $\gamma$, $\Omega$, and $k_BT$ as in Fig.~\ref{case5}, we have found strong dependence on the initial conditions of the positivity violation. 
	In Fig.~\ref{fig:zeta_phi}, a large dark region corresponds to the complex $\zeta$'s for which positivity violations can happen for the Markovian runs. This is further supported in Fig.\ref{case25}, where 
	two individual time evolutions are shown with the same $|\zeta|$, but opposite sign of $\phi$. For $\phi=-\pi/4$, the quotient $A/C$ shows a strong positivity violation, namely, in a small-time interval it becomes 
	negative. There is no violation for $\phi=\pi/4$. This particular 
	situation
	is explained by inequality \eqref{eq:jolt_mar} at $t=0$ (see Appendix \ref{sec:initial_jolt}). In fact, $c_2(0)$ flips sign for the change $\phi\to-\phi$.
	In the non-Markovian case, we found no positivity violations at all
	for this family of initial conditions if the stationary solution is physical.

	Next, we discuss what can happen if one starts from a thermal state (which is not a pure initial state for $T'>0$). Let us consider Fig.~\ref{thermal_min}. We plot the minimal values of the quotient $A/C$ for 
	individual Markovian runs starting from thermal initial states. Different curves belong to different damping factors $\gamma$. At $T'=0$, we start from a coherent state. All relevant parameters belong to region II. 
	The figure clearly supports the expectation that if one increases the width of the initial Gaussian, one can avoid positivity violations. Curves with decreasing $\gamma$ are further away from the critical line. 
	Choosing $\gamma$ to be bigger than $0.72$, there is no positivity violation even for $T'=0$.

	These numerical investigations suggest that the non-Markovian evolution becomes unphysical, i.e., $A/C <1$, only when the stationary state is unphysical. This has been investigated in detail in Sec. \ref{III} 
	and results in constraints on the choice of the parameters of the model. However, this is not true for the Markovian evolution, which may show, for certain times of the evolution unphysical behavior. 
	It is indeed true 
	that the non-Markovian evolution is still more reliable  than the Markovian one.

	\section{Summary and final remarks}
	\label{V}
	
	Summarizing, we have investigated a HPZ master equation of the Caldeira-Leggett model with a quantum harmonic oscillator, where we have considered the weak-coupling limit up to the second-order 
	in the coupling parameter and Ohmic spectral density with a Lorentz-Drude cutoff function. The restriction to weak-coupling does not necessarily mean  that the influence of the bath on the system is 
	weak, i.e., weak damping. The large number of bath modes may act collectively and thereby have a strong influence on the open system even if each mode is perturbatively weakly coupled to it;
	see, for example \cite{GT}. Therefore, we have begun our analysis without any restriction on the parameters of model.
	
	Our goal has been to identify unphysical behavior of this master equation by means of following time evolutions of the initial density operators and examining whether the 
	evolving density operators lose their positivity. This is a very delicate problem for general initial density operators, because the time evolution is usually followed in the phase-space representation and the 
	study of positivity properties of the Weyl transformed operators is still an open problem \cite{Nicola}. 
	Therefore, we have focused only on Gaussian states, where the spectrum can be completely identified from the phase-space solutions of the master equation. 
	
	As a first step, in Sec. \ref{II}, we have transformed the whole problem into a phase-space representation where the evolution is described by a linear differential equation system. Then, we have identified 
	algebraic relations between the evolving coefficients of this phase-space representation and the spectrum of the evolving operator, which may not always be a density operator. We have used numerical 
	simulations to follow the evolving spectrum. We have compared the non-Markovian evolution to a Markovian one, which we have obtained by taking the  coefficients in the $t \to \infty$ limit; see 
	Eq.~\eqref{diffcoeffM}. We have showed for coherent, squeezed and thermal initial conditions that the positivity violations in the non-Markovian evolution occur when the stationary solution is also no longer a physical 
	state. Therefore, a positivity check on the stationary solution is necessary, which puts important constraints on the parameters of our theory. Therefore, we have carried out an analysis on the 
	stationary solution in Sec. \ref{III}, where we have also found results known by the community, see \cite{CLLT} or \cite{Lampo2}. However, it is worthwhile to mention that not all published material 
	handles this positivity issue very carefully; see, for example, Fig. $10.7$ in \cite{book1}. In contrast to the non-Markovian evolution, we have found in Sec. \ref{IV}, both for short (occurring at $t=0^+$) and 
	intermediate (occurring at finite $t>0$) time evolutions, positivity violations in the Markovian case. Our numerical investigations suggest that the rapid growth of the diffusion coefficient $D_{pp}(t)$ compared to the 
	growth of $D_{px}(t)$ is the reason, why the non-Markovian master equation avoids positivity violations for short evolution times. 
	
	We have only considered Ohmic spectral density with a Lorentz-Drude cutoff function, but one may ask 
	what can happen for other types of spectral densities. At least we know from \cite{Hu_Paz} that in cases of so-called supra- and subohmic spectral densities, $D_{pp}(t)$ is growing faster than $D_{px}(t)$ for short 
	times and, together with our results, we conjecture that non-Markovian evolutions for these spectral densities also cannot exhibit positivity violations for Gaussian initial states and physical
	stationary states.
	
	If one considers the time evolution (\ref{NMEQ}) starting from an \emph{arbitrary}, not necessarily Gaussian, initial density operator, then one can state the following: for parameters belonging to  
	region I of Fig.~\ref{density} and starting from any initial condition, there must be positivity violation both for non-Markovian and Markovian master equations. This can be explained as follows. 
	For parameters in region I the asymptotic state is non-physical. However, this state is unique and corresponds to the asymptotic Gaussian state of any initially physical state, e.g., 
	see \cite{Fleming,Lampo2} discussed in their Sec. III. If this state is non-physical, then positivity violation must occur at least asymptotically. For parameters in regions II and III
	one should not rule out the possibility of finding positivity violations for appropriately chosen general initial density operators as in the case of Gaussian initial states and the Markovian master equation.
	
	Numerically, the non-Markovian evolution does not seem to show any signs of positivity violations for physical stationary states. Unfortunately, this is not always the case for the 
	Markovian evolution. Therefore, we may say the non-Markovian evolution is superior to the Markovian one, which is, vaguely speaking due  to the rapid growth of $D_{pp}(t)$ compared to that of $D_{px}(t)$. 
	We managed to prove  in Appendix~\ref{sec:initial_jolt}  that there is no short time positivity violation for the arbitrary Gaussian initial state and parameters of the model. 
	This remains true even when the stationary solution is unphysical.  This finding seems to be connected to the so-called initial ``jolt'' found by Refs. \cite{Unruh,Hu_Paz}.
	
	A few generic comments on the purity of the evolving solutions are in order. In our whole investigation, we have focused on the ratio $A/C$ which, in turn, is the squared inverse of the purity.
	Thus, all figures implicitly describe  the purity as well, which is a measure of mixedness. Many figures show that purities are non monotonic in time and therefore states undergo a certain amount of 
	purification or mixing during the time evolution. An easy way to understand this effec is to consider an initial pure state and a different pure stationary state. As the dynamic is clearly not unitary, 
	the stationary state will be reached throughout not necessarily pure states and thus purity in this example cannot be monotonic; see our Fig. \ref{case11}. 
	
	Several questions concerning this subject remain  open problems, even though applications of these master equations are very frequent. Here, we have thoroughly investigated 
	a Markovian and a non-Markovian master equation of the Caldeira-Leggett model for initial Gaussian density operators and identified the boundaries of the physically interpretable solutions of 
	the time evolutions. Therefore, our results provide a key step in establishing the range of applicability of these master equations.
	
	\section*{Acknowledgement}
	
	The authors have profited from helpful discussions
	with  M. Kornyik, L. Lisztes,  G. Helesfai,  Z. Kaufmann and \'E. Valk\'o. This research is supported by the National Research
	Development and Innovation Office of Hungary within
	the Quantum Technology National Excellence Program
	(Project No. 2017-1.2.1-NKP-2017-00001) and the European Union’s Horizon 2020 research and innovation
	programme under Grant Agreement No. 732894 (FET
	Proactive HOT). M.A.C. was supported by the Norwegian Research Council through  Grants
	No. 287906 and No. 262695 (CoE Hylleraas Centre for Quantum
	Molecular Sciences).  M.A.C. also received support of the NKFIH through the National Quantum
	Technology Program (Grant No. 2017-1.2.1-NKP-2017-
	00001) and Grant No. K120569.

\appendix
\section{Expressions for the coefficients \texorpdfstring{$D_{pp}(t)$}{TEXT} and \texorpdfstring{$D_{px}(t)$}{TEXT}}
\label{Appuseful}
	\label{Appuseful}
Expanding the $\coth$ function in eq. (\ref{eq:D1(s)_def}) as
\[
\coth \pi x=\sum_{n=-\infty}^\infty \frac{x}{\pi(x^2+n^2)}=\frac{1}{\pi x}+\frac{2x}{\pi}\sum_{n=1}^\infty \frac{1}{(x^2+n^2)}
\]
and integrating term by term one gets for 
$s>0$ 
%eq.(\ref{d1asasum}) can be written as
\begin{eqnarray}
D_1(s)&=& 
\frac{4 m \gamma k_B T \Omega^2}{\hbar}%\times\nonumber \\ &&
\left[\frac{e^{-\Omega s}}{\Omega} +2 \sum_{n=1}^{\infty}\frac{\Omega e^{-\Omega s}- \nu_n e^{-\nu_n s} }{\Omega^2-\nu_n^2} \right], \nonumber\\
\label{eq:d1sumshort}
\end{eqnarray}
where $\nu_n$'s are the bosonic Matsubara frequencies:
\begin{equation}
\nu_n=2\pi n k_B T/\hbar.
\end{equation}

The first part in the square brackets of (\ref{eq:d1sumshort}) can be transformed using the identity
\[
e^{-\Omega s}\left[\frac{1}{\Omega}+2\sum_{n=1}^\infty \frac{\Omega}{\Omega^2-\nu_n^2}\right]=\frac{\pi}{\nu_1}\cot\left( \frac{\Omega \pi}{\nu_1}\right)e^{-\Omega s},
\]
where $\nu_1=2\pi k_BT/\hbar$ is the first bosonic Matsubara-frequency.
The other part in the square brackets of (\ref{eq:d1sumshort}) can be expressed as
\begin{eqnarray}
&&\sum_{n=1}^{\infty}\frac{ \nu_n e^{-\nu_n s} }{\Omega^2-\nu_n^2}=-\frac{e^{-\nu_1 s}}{2\nu_1}
\times\nonumber \\
\!\!\!\!&&\!\!\!\!\!\left( G\left(e^{-\nu_1 s},1,1-\frac{\Omega}{\nu_1} \right)+G\left(e^{-\nu_1 s},1,1+\frac{\Omega}{\nu_1} \right)\right), 
\end{eqnarray}
where $G(z,a,b)$ denotes the so-called Lerch transcendent or  HurwitzLerchPhi$[z, a, b]$ in Mathematica \cite{hurwitzlerchpi}.

A similar but different sum also appear later
\begin{eqnarray}
&&\sum_{n=1}^{\infty}\frac{ \nu_n e^{-\nu_n s} }{\omega_0^2+\nu_n^2}= \nonumber\\
\!\!\!\!&&\!\!\!\!\! \frac{ e^{-i\omega_0 s}F\left(e^{-\nu_1 s},\frac{\nu_1-i\omega_0}{\nu_1},-1 \right)+e^{i\omega_0 s}F\left(e^{-\nu_1 s},\frac{\nu_1+i\omega_0}{\nu_1},-1 \right)}{2i\omega_0}, 
\nonumber
\end{eqnarray}
where $F(z,a,b)$ is the so-called incomplete beta function Beta$[z,a,b]$ (we also use the terminology of Wolfram Mathematica).

We need also two more sums over the Matsubara-frequencies, however, those ones can be calculated  via the useful formulas
\[
\sum_{n=1}^{\infty}\frac{ \nu_n^2 e^{-\nu_n s} }{\Omega^2-\nu_n^2}=-\frac{\partial}{\partial s} \left( \sum_{n=1}^{\infty}\frac{ \nu_n e^{-\nu_n s} }{\Omega^2-\nu_n^2}\right),
\] 
\[
\sum_{n=1}^{\infty}\frac{ \nu_n^2 e^{-\nu_n s} }{\omega_0^2+\nu_n^2}=-\frac{\partial}{\partial s} \left( \sum_{n=1}^{\infty}\frac{ \nu_n e^{-\nu_n s} }{\omega_0^2+\nu_n^2}\right).
\] 
Inserting the series (\ref{eq:d1sumshort}) into eq.(\ref{diffcoeff}) the integral over $s$ is trivial, but the final forms for the diffusion coefficients are lengthy:
\begin{widetext}
	\begin{eqnarray}	
	D^{(2)}_{px}(t)=\frac{k_B T \gamma\Omega^2}{\hbar \omega_0\left( \omega^2_0+\Omega^2\right) }
	\Biggl\{ &&
	\omega_0 \cdot \left( \frac{1}{\Omega}+ 2\sum_{n=1}^\infty \frac{\Omega}{\Omega^2-\nu_n^2}- 2\sum_{n=1}^\infty \left[ \frac{ \nu_n }{\Omega^2-\nu_n^2}+ \frac{ \nu_n}{\omega_0^2+\nu_n^2}\right] \right) \Biggr. \nonumber \\
	&&-\omega_0\cos{(\omega_0 t)} \cdot	\left(\frac{e^{-\Omega t}}{\Omega}+ 2\sum_{n=1}^\infty \frac{\Omega e^{-\Omega t}}{\Omega^2-\nu_n^2}- 2\sum_{n=1}^\infty \left[ \frac{ \nu_n e^{-\nu_n t}}{\Omega^2-\nu_n^2}+ \frac{ \nu_n e^{-\nu_n t}}{\omega_0^2+\nu_n^2}\right]\right) \nonumber \\
	&&	\Biggl. -\sin{(\omega_0 t)} \cdot\left(  e^{-\Omega t}+ 2\sum_{n=1}^\infty \frac{\Omega^2 e^{-\Omega t}}{\Omega^2-\nu_n^2}- 2\sum_{n=1}^\infty \left[ \frac{ \nu_n^2 e^{-\nu_n t}}{\Omega^2-\nu_n^2}+ \frac{ \nu_n^2 e^{-\nu_n t}}{\omega_0^2+\nu_n^2}\right]\right) \Biggr\} \label{eq:D_px(2)}
	\end{eqnarray}
	\begin{eqnarray}	
	D^{(2)}_{pp}(t)=\frac{2k_B T m\gamma\Omega^2}{\hbar\left( \omega^2_0+\Omega^2\right) }
	\Biggl\{ &&
	\left( 1 + 2\sum_{n=1}^\infty \frac{\Omega^2}{\Omega^2-\nu_n^2}- 2\sum_{n=1}^\infty \left[ \frac{ \nu_n^2 }{\Omega^2-\nu_n^2}+ \frac{ \nu_n^2}{\omega_0^2+\nu_n^2}\right] \right) \Biggr. \nonumber \\
	&&+\omega_0\sin{(\omega_0 t)} \cdot	\left(\frac{e^{-\Omega t}}{\Omega}+ 2\sum_{n=1}^\infty \frac{\Omega e^{-\Omega t}}{\Omega^2-\nu_n^2}- 2\sum_{n=1}^\infty \left[ \frac{ \nu_n e^{-\nu_n t}}{\Omega^2-\nu_n^2}+ \frac{ \nu_n e^{-\nu_n t}}{\omega_0^2+\nu_n^2}\right]\right) \nonumber \\
	&&	\Biggl. -\cos{(\omega_0 t)} \cdot\left(  e^{-\Omega t}+ 2\sum_{n=1}^\infty \frac{\Omega^2 e^{-\Omega t}}{\Omega^2-\nu_n^2}- 2\sum_{n=1}^\infty \left[ \frac{ \nu_n^2 e^{-\nu_n t}}{\Omega^2-\nu_n^2}+ \frac{ \nu_n^2 e^{-\nu_n t}}{\omega_0^2+\nu_n^2}\right]\right) \Biggr\} \label{eq:D_pp(2)}.
	\end{eqnarray}
	
\end{widetext}
We used the above formulas in our numerical works. The Markovian values for $\omega_p^2$ and $\lambda$ are
\begin{equation}
\left(\omega_p^{(M)}\right)^2=\omega^2_0+2 \gamma \Omega-\frac{2\gamma \Omega^3}{\Omega^2+\omega^2_0}, \quad \lambda^{(M)}=\frac{\gamma\Omega^2}{\Omega^2+\omega_0^2}.
\label{eq:Markovian_lambda_omega}
\end{equation}
The asymptotic Markovian values
for the diffusion coefficients can be read off from the first lines of eqs. (\ref{eq:D_px(2)}) and (\ref{eq:D_pp(2)}). Performing the Matsubara sums they can be given as
\begin{equation}
D_{pp}^{(M)}=m\gamma \omega_0 \frac{\Omega^2}{ \omega^2_0+\Omega^2}\coth\left(\frac{\hbar\omega_0} {2k_BT}\right),
\end{equation}
\begin{eqnarray}
D_{px}^{(M)}&=&\frac{\gamma\Omega^2}{\Omega^2+\omega_0^2}\Biggl[ -\frac{k_B T}{\hbar\Omega}-\frac{1}{2\pi}\biggl\{2\Psi\left( \frac{\hbar\Omega}{2\pi k_B T}\right)    \biggr.\Biggr. \nonumber\\
&& \Biggl.\biggl. \quad-\Psi\left( \frac{i\hbar\omega_0}{2\pi k_B T} \right)-\Psi\left( \frac{-i\hbar\omega_0}{2\pi k_B T} \right)
\biggr\} \Biggr],
\label{eq:Markovian_Dpx}
\end{eqnarray}
where $\Psi(x)$ is the digamma function. The Markovian values \eqref{eq:Markovian_lambda_omega}-\eqref{eq:Markovian_Dpx} fully determine the asymptotic matrix $\mathbf{M}^{(M)}$ and the asymptotic vector $\mathbf{v}^{(M)}$.

\section{Behavior of \texorpdfstring{$D_{pp}(t)$}{TEXT} and  \texorpdfstring{$D_{px}(t)$}{TEXT} for small time \texorpdfstring{$t$}{TEXT}}
\label{sec:dpp_app}

	At very small temperature the hyperbolic cotangent factor in Eq.(\ref{eq:D1(s)_def}) can be well approximated by one:
\begin{eqnarray}
&&D_1(s)|_{T=0} = \frac{2\gamma m \Omega^2}{\pi} \cdot\int_0^\infty \frac{\omega}{\Omega^2+\omega^2}\cos(\omega s) d\omega =  \nonumber \\
&&= \frac{2\gamma m \Omega^2}{\pi} \Bigl(\sinh(\Omega s)\textrm{Shi}\,(\Omega s)-\cosh(\Omega s)\textrm{Chi}\,(\Omega s) \Bigr), \qquad\label{eq:d1_at_T_eq_0}
\end{eqnarray}
where 
\begin{equation}
\textrm{Chi}\,(z)=\gamma_{\mathrm{EM}}+\ln(z)+\int_0^z\frac{(\cosh(t)-1)}{t}dt,
\label{eq:chi}
\end{equation}
is the function CoshIntegral[x] and 
\begin{equation}
\textrm{Shi}\,(z)=\int_0^z \frac{\sinh(t)}{t}dt
\label{eq:shi}
\end{equation}
is the function SinhIntegral[x] in Mathematica. For short times $s$ the dominant behavior in $D_1(s)$ is the logarithm function. By Eqs.~(\ref{diffcoeff}), (\ref{eq:chi}) and (\ref{eq:shi}) the coefficients  $D_{pp}(t)$ and $D_{px}(t)$ behave as
\begin{equation}\label{eq:Dpplead}
D_{pp}(t) = \frac{2\gamma m\Omega^2}{\pi}\left(1-\gamma_{\mathrm{EM}}-\ln\Omega t \right)t+\mathcal{O}(t^3),
\end{equation} 
\begin{equation}\label{eq:Dpxlead}
D_{px}(t) = \frac{\gamma \Omega^2}{4\pi}\left(1-2\gamma_{\mathrm{EM}}-2\ln\Omega t \right)t^2+\mathcal{O}(t^4).
\end{equation} 
for small $t$ and  $T=0$. 

At finite temperature one can make the decomposition
\begin{eqnarray}
&&D_1(s)=D_1(s)|_{T=0} \nonumber \\
&&+\frac{2\gamma m \Omega^2}{\pi} \int_0^\infty \frac{\omega}{\Omega^2+\omega^2}\left[
\coth\left(\frac{\hbar\omega}{2k_BT}\right)-1
\right]d\omega, \nonumber
\end{eqnarray}
where the first term  on the right hand side is discussed above and behaves as $\sim\ln(\Omega s)$, while the second is finite even for $s=0$. By Eq. (\ref{diffcoeff}) at finite temperature the short time dominant behavior of $D_{px}(t)$ and $D_{pp}(t)$ are still:
\begin{eqnarray}
D_{pp}(t) &\simeq& -\frac{2\gamma m\Omega^2}{\pi}t \ln(\Omega t), \label{eq:Dpptruelead} \\
D_{px}(t) &\simeq& -\frac{\gamma\Omega^2}{2\pi}t^2 \ln(\Omega t). \label{eq:Dpxtruelead}
\end{eqnarray}

	\section{Analysis of small-time behavior}
\label{sec:initial_jolt}

In this appendix, we show how a differential equation for the quotient $A(t)/C(t)$ can be used to prove 
small-time positivity violation/non-violation. We begin with the non-Markovian case. Using the notations of Section \ref{III}, we set 
$Q(t)=A(t)/C(t)=16c_1(t)c_3(t)-4c_2^2(t)$, and via the system \eqref{eq:main_diffeq} we arrive at
\[
\dot{Q}+4\lambda(t)Q=16 \frac{D_{pp}(t)}{\hbar}c_1(t) - 16D_{px}(t)c_2(t).
\]
The general solution of which is given by the variation of constants formula
\begin{eqnarray}
&&Q(t)=\frac{Q(0)}{\Lambda(t)}+  \nonumber\\ 
&&+\frac{16}{\Lambda(t)} \int_0^t \Lambda(s) \biggl[\frac{D_{pp}(s)}{\hbar}c_1(s)-D_{px}(s)c_2(s)\biggr] \, ds, \nonumber
\end{eqnarray}
where we have let $\Lambda(t)= \exp\left(4\int_0^t\lambda\right)$ for convenience.
Using this, the condition $Q(t)\ge 1$ is clearly equivalent to $F(t)\ge 0$, where
\begin{eqnarray}
F(t)&=&\int_0^t \Lambda(s)\biggl[\frac{D_{pp}(s)}{\hbar}(s)c_1(s)-D_{px}(s)c_2(s)\biggr] \, ds \nonumber \\
&& - \frac{\Lambda(t)-Q(0)}{16}.\nonumber
\end{eqnarray}
Note that $F(0)=\frac{Q(0)-1}{16}\ge 0$.
Therefore, a sufficient condition for $F(t)\ge 0$ for small $t$ to hold is simply that
$F'(t)\ge 0$, i.e.,
\begin{equation}
\frac{D_{pp}(t)}{\hbar}c_1(t)-D_{px}(t)c_2(t)\ge \frac{\lambda(t)}{4}.
\label{eq:jolt_eq}
\end{equation}
We note in passing that for pure initial states, $F(0)=0$, so $F'(t)\ge 0$ is actually equivalent to $Q(t)\ge 1$ for sufficiently small $t$.
Using the expressions \eqref{eq:Dpptruelead} and \eqref{eq:Dpxtruelead}  and the short time dominant behavior  $\lambda(t)\simeq \tfrac{1}{2}\gamma\Omega^2t^2$, we find
\begin{equation}
-\frac{2m}{\pi\hbar}\ln(\Omega t)\cdot c_1(t)+ \frac{1}{2\pi}t\ln(\Omega t) \cdot c_2(t) \gtrsim \frac{t}{8},\nonumber
\end{equation}
which is obviously true for any trajectory $(c_1,c_2,c_3)$ for sufficiently small $t$. 
In fact, $c_1$ is always positive, and on the left hand side the first term is bigger in modulus than the second term. In the first term the logarithm ensures that the inequality is true for small $t$, and for any positive $c_1$.
This shows that
the non-Markovian time evolution never violates positivity at $t=0^+$.

In the Markovian case, a completely analogous condition to \eqref{eq:jolt_eq} can be derived with $D_{pp}(t)$, $D_{px}(t)$ and $\lambda(t)$
replaced by their Markovian counterparts $D_{pp}^{(M)}$, $D_{px}^{(M)}$ and $\lambda^{(M)}$, viz.
\begin{equation}\label{eq:jolt_mar}
\frac{D_{pp}^{(M)}}{\hbar}c_1(t)-D_{px}^{(M)}c_2(t)\ge \frac{\lambda^{(M)}}{4}.
\end{equation}
Now consider squeezed initial states $\mathbf{c}_\mathrm{sq}(0)$, for which clearly $Q(0)=1$. Evaluating the preceding inequality at $t=0$,
we obtain a set of initial states $\mathbf{c}_\mathrm{sq}(0)$ that is surely violating at $t=0^+$. This constitutes a subset of
the gray set in Figure \ref{fig:zeta_phi}. Hence, we have shown that in the Markovian case, it is always possible to find a pure state that violates positivity at $t=0^+$.

\end{document}